\documentclass[aps,pra,reprint,amsmath,showpacs]{revtex4-1}

\usepackage[utf8]{inputenc}  
\usepackage[ngerman,american]{babel}
\usepackage{times}
\usepackage[T1]{fontenc}
\usepackage{amsmath, amssymb,amsfonts,mathtools}
\usepackage{graphicx}

\usepackage[normalem]{ulem}

\usepackage{braket}

\usepackage{hyperref}

\usepackage{color}
\newcommand{\vg}{v_{\mathrm g}}
\newcommand{\La}{L_\mathrm{abs}}
\newcommand{\Lc}{L_\mathrm{corr}}

\newcommand{\bd}{b^\dagger}
\newcommand{\Hrm}{\hat{\mathrm H}}
\newcommand{\Psid}{\hat\Psi^\dagger}
\newcommand{\Psih}{\hat\Psi}
\newcommand{\Phih}{\hat\Phi}
\newcommand{\Phid}{\hat\Phi^\dagger}

\newcommand{\E}{\hat{\mathcal{E}}}
\newcommand{\Ed}{\hat{\mathcal{E}}^\dagger}

\newcommand{\ke}{\ket{e}}
\newcommand{\kr}{\ket{r}}
\newcommand{\kg}{\ket{g}}

\newcommand{\sge}{\hat\sigma_\mathrm{ge}}

\newcommand{\sgr}{\hat\sigma_\mathrm{gr}}

\newcommand{\see}{\hat\sigma_\mathrm{ee}}
\newcommand{\srr}{\hat\sigma_\mathrm{rr}}

\newcommand{\intdz}{\int\!\mathrm dz}

\newcommand{\pdt}{\partial_t}
\newcommand{\pdz}{\partial_z}

\newcommand{\ODB}{\mathrm{OD}_\mathrm B}
\newcommand{\OD}{\mathrm{OD}}

\newcommand{\abs}[1]{\vert #1 \rvert}
\newcommand{\mat}[2]{\begin{pmatrix}#1\\#2\end{pmatrix}}

\renewcommand{\vec}{\mathbf}
\usepackage{import}

\begin{document}

\title{Many-body physics of Rydberg dark-state polaritons in the strongly interacting regime}
\author{Matthias Moos, Michael H\"oning, Razmik Unanyan, and Michael Fleischhauer}
\affiliation{Fachbereich Physik und Forschungszentrum OPTIMAS,
Technische Universit\"at Kaiserslautern, 67663 Kaiserslautern, Germany}
\date{\today}

\begin{abstract}
Coupling light to Rydberg states of atoms under conditions of electromagnetically induced transparency (EIT) leads to the formation of strongly interacting quasi-particles, termed Rydberg polaritons. We derive a one-dimensional model describing the time evolution of these polaritons under paraxial propagation conditions, which we verify by numerical two-excitation simulations. We determine conditions allowing for a description by an effective Hamiltonian of a single-species polariton, and calculate ground-state correlations by use of the density matrix renormalization group (DMRG). Under typical stationary slow-light EIT conditions it is difficult to reach the strongly interacting regime where the interaction energy dominates the kinetic energy. We show that by employing time dependence of the control field the regime of strong interactions can be reached where the polaritons attain quasi crystalline order. We analyze the dynamics and resulting correlations for a translational invariant system in terms of a time-dependent Luttinger liquid theory and exact few-particle simulations and address the effects of nonadiabatic corrections and initial excitations.
\end{abstract}


\maketitle
\section{Introduction}
Recently there has been growing interest in using Rydberg gases~\cite{Gallagher1994} to tailor strong and nonlocal nonlinearities for photons. The strong interactions between 
Rydberg states in a gas of ultra-cold atoms can mediate strong nonlocal interactions for light fields propagating under conditions of electromagnetically induced transparency (EIT) in such a medium~\cite{Bason2008,Pritchard2010,Sevincli2011,Petrosyan2011,Olmos2011,Honer2011,Hofmann2013,Maxwell2014,Maghrebi2015}. For instance, the formation of a small blockaded volume leading to antibunching of photons as well as pronounced bunching of photons in a Rydberg gas have been demonstrated experimentally~\cite{Peyronel2012,Firstenberg2013}. Moreover, first steps have been taken towards building single photon logic devices, such as single photon switches~\cite{Baur2014,Tiarks2014,Gorniaczyk2014,Paredes-Barato2014}. 

Rydberg polaritons exhibit potentially strong and nonlocal mutual interactions making them interesting candidates to study many-body effects in the regime of strong correlations, where the interaction energy dominates the kinetic energy. 
In the present paper we consider the propagation of light in Rydberg EIT media under paraxial conditions in a one-dimensional setting, as can be realized experimentally by, e.g., cigar-shaped atomic ensembles or sending photons through hollow-core optical fibers filled with Rydberg atoms~\cite{Shahmoon2011,Epple2014}. In a previous letter~\cite{Otterbach2013} we argued that the propagation of photons inside Rydberg gases under EIT conditions can be described in terms of interacting quasi-particles, named Rydberg polaritons. We showed that a regime of strong correlations, where the ratio between interaction energy and kinetic energy becomes large, can be reached by dynamically turning photons into stationary Rydberg excitations~\cite{Maxwell2013}.

In the present paper we extend these studies by deriving an effective dark-state polariton model including leading-order corrections valid for sufficiently large separation of Rydberg polaritons. This regime allows for a perturbative treatment of the coupling between bright- and dark-state polaritons. We derive a Lindblad master equation for the Rydberg dark-state polaritons where the bright-state polaritons act as a Markovian reservoir. We verify the model and address the preparation of an initial state by employing numerical two-excitation wave-function simulations. We derive conditions for when the Master equation can be reduced to the effective Hamiltonian introduced in~\cite{Otterbach2013}. Furthermore, we show that for an excitation density smaller than one per blockade volume and a transversal beam diameter less than the blockade radius the paraxial propagation can be described by a one-dimensional model. 

\begin{figure}
 \centering
\includegraphics[width=0.8\columnwidth]{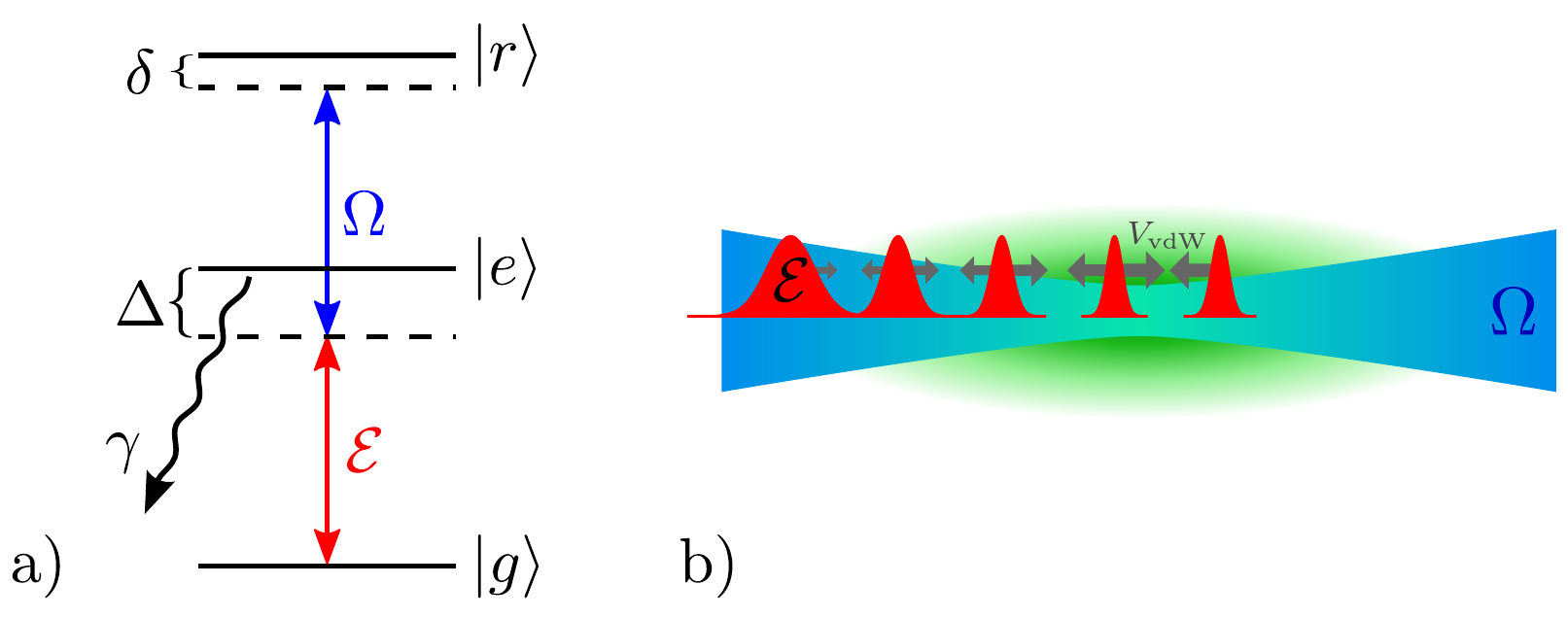}
 \caption{%
 a) Atomic level scheme with input probe field $\mathcal E$ and control field $\Omega$.
 b) Sketch of a possible experimental realization: one dimensional setup with copropagating photons focused inside an elongated atomic cloud of Rydberg atoms and control field $\Omega$.
 }%
 \label{fig_level_scheme}
\end{figure}

The paper is organized as follows. In Section II we introduce the microscopic model of photons coupled to interacting three-level atoms. We map the model to the polaritonic basis and derive a one-dimensional Born-Markovian master equation for the dark-state polaritons by treating the bright-state polaritons as a reservoir.
Calculating interaction-mediated scattering amplitudes between different transverse modes we derive conditions for an effective one-dimensional treatment of the problem. 
We furthermore analyze conditions when the master equation reduces to an effective Hamiltonian describing a unitary time evolution. In section III we employ numerical two-excitation wave-function calculations to verify the model and determine the initial state after sending a light pulse into a Rydberg gas. In section IV we calculate ground state correlations of the effective Hamiltonian by use of density-matrix renormalization group (DMRG) calculations and a Luttinger liquid approach. We show that under continuous driving conditions and large optical depth per blockade distance the regime of strong interactions cannot be entered since the Rydberg blockade prevents to reach the polariton densities necessary for the interaction energy to overcome kinetic contributions. However,  in section V we show that it is possible to enter this regime by dynamically slowing down Rydberg polaritons. In particular we consider the storage of interacting photons inside a Rydberg medium. We apply numerical wave function simulations to show that the losses during storage remain finite and use a time-dependent Luttinger liquid approach to compute time-dependent density correlation functions.

\section{Rydberg polaritons}

In this section, we discuss the paraxial propagation of weak light pulses consisting of few photons in an atomic medium under EIT conditions with Rydberg interactions.

\subsection{Light-matter coupling}
%
%

Let us consider a system of a weak quantized probe field $\hat E=\sqrt{\frac{\hbar\omega_p}{2\epsilon_0}}\E e^{-i(\omega_p t-k_pz)}+\mathrm{h.a.}$ propagating through an ensemble of  non-interacting three level atoms as sketched in Fig.~\ref{fig_level_scheme} b) with the level structure in ladder configuration as in~\cite{Fleischhauer2000}, and as drawn in Fig.~\ref{fig_level_scheme} a).
The operators $\E,\Ed$ are slowly varying envelope operators, which obey bosonic commutation relations $[\E(\vec r),\Ed(\vec r')]=\delta(\vec r-\vec r')$ and $\omega_p$ and $k_p$  denote carrier frequency and wave vector of the probe field, respectively. The atoms are composed of a ground state $\ket g$, an intermediate excited state $\ket e$ and a metastable Rydberg state $\ket r$. Here we neglect dipole-dipole interactions of Rydberg states for a moment, which we will reintroduce later. The field $\E$ is coupled to the  coherence between atomic states $\kg$ and $\ke$ with single-photon detuning $\Delta=\omega_{eg}-\omega_\mathrm{p}$. The coherence between states $\ke$ and $\kr$ is driven by a classical control field $\Omega$ with carrier frequency $\omega_c$ and detuning $\Delta_c=\omega_\mathrm{re}-\omega_c$. We denote the resulting two-photon detuning by $\delta=\Delta+\Delta_c$. The atomic state $\ke$ is assumed to be subject to spontaneous decay with rate $\gamma$.

The atomic polarization is microscopically described by spin flip operators $\ket\mu\bra\nu$ of individual atoms.  By averaging these over a small volume centered at a position $\vec r$ and containing $N_\vec r\gg 1$ atoms we define continuous, coarse-grained atomic spin flip operators $\sigma_{\mu\nu}$ at position $\vec r$,
\begin{equation}\label{eq:coarse_graining}
\hat\sigma_{\mu\nu}(\vec r) = \frac{1}{N_\vec r}\sum_{j=1}^{N_\vec{r}}\ket{\mu}_{j\,j}\!\bra\nu,
\end{equation}
which  fulfill the commutation relations
%
%
$\left[{\hat\sigma}_{\alpha\beta}(\vec r),{\hat\sigma}_{\mu\nu}(\vec r')\right]=\frac{1}{n}\delta(\vec r-\vec r')\left[\delta_{\beta\mu}{\hat\sigma}_{\alpha\nu}(\vec r)-\delta_{\alpha\nu}{\hat\sigma}_{\mu\beta}(\vec r)\right],
$
%
%
where the density $n$ of the atoms is assumed to be homogeneous.
Transforming to a frame rotating with the atomic frequencies 
and performing the rotating wave approximation, the atom-light coupling Hamiltonian of this system can be written as
($\hbar = 1$)
\begin{multline}\label{eq:H_light_atom}
\quad\Hrm =n\int {\mathrm d}^3\mathbf{r}\ \Bigl\{\Delta\hat\sigma_{ee}(\vec r)+\delta\hat\sigma_{rr}(\vec r)\Bigr\}\\
-\Bigl\{\Omega\hat\sigma_{re}(\vec r)
+g\sqrt{n}\hat{\mathcal E}(\vec r)\hat\sigma_{eg}(\vec r)+\mathrm{h.a.}\Bigr\},\quad
\end{multline}
where $g=d_{ge}\sqrt{\omega_{ge}/2\hbar\epsilon_0}$ denotes the coupling strength of the electric field $\E$ 
to the atomic transition $\kg-\ke$, with $d_{ge}$ being the atomic dipole matrix element.

We derive Heisenberg-Langevin equations of motion for the slowly varying field $\E$ and atomic operators taking into account the spontaneous decay rate $\gamma$ of the intermediate level~\cite{Louisell1973}. Assuming the probe field to be weak compared to the control field we can treat the equations in linear response with respect to $g\E$, leading to 
%
%
\begin{equation}\label{eq:Heisenberg_Langevin_coherences}
\begin{aligned}
\partial_t\sgr& =-i\delta\sgr+ i\Omega\sge\\
\partial_t\sge& =-\Gamma\sge+ig\sqrt{n}\E+ i\Omega\sgr+\hat F_{ge},
\end{aligned}
\end{equation}
where $\Gamma=\gamma+i\Delta$. $\hat F_{ge}$ denotes a Langevin noise operator~\cite{Louisell1973}, which is $\delta$-correlated in time and space with vanishing expectation value that needs to be added to preserve the commutation relations. As the noise is related to the population of the excited state which we set $ \see=0 $ in linear response, the Langevin operators can be neglected.

In the following we want to consider the case of two-photon resonant driving, i.e., $\delta=0$. We note, however, that taking interactions between atoms excited to level $\ket r$ into account will lead to an effective space-dependent two photon detuning.

\subsection{Paraxial light propagation}
\label{subsec:paraxial_light_propagation}
%
The dynamics of the probe field is described by the truncated paraxial wave equation,
\begin{equation}\label{eq:truncated_wave_equation}
\left(\frac{\partial}{\partial t}+c\frac{\partial}{\partial z}-i\frac{c}{2k_p}\nabla_\perp^2\right)\E(\vec r,t)=ig\sqrt{n}{\hat\sigma}_{ge}(\vec r,t).
\end{equation}
We assume a cylindrical symmetry of the setup and decompose the probe field into mode functions $u_{kl}(r,\varphi)$, which are eigensolutions of $\nabla_\perp^2u_{kl}(r,\varphi)=0$,
\begin{equation}
\E(\vec r,t)=\sum_{k,l}u_{k,l}(r,\varphi)\E_{k,l}(z,t).
\label{eq:decompose}
\end{equation}
The mode functions
\begin{equation}
u_{kl}(r,\varphi)=\frac{C_{kl}}{w_0}\left[\frac{\sqrt 2 r}{w_0}\right]^{\!\abs{k}}\!\!e^{-r^2/w_0^2+ik\varphi}L_l^{|k|}\left(\frac{2r^2}{w_0^2}\right)\!,
\end{equation}
are complete orthogonal set in $(r,\phi)$.  $ L_l^{k} $ are the associated Laguerre polynomials and $ C_{kl} $ are appropriate normalization constants. $l$ and $k$ denote the  radial and azimuthal index of the mode functions, respectively.
The decomposition (\ref{eq:decompose}) is adequate as it 
describes Gauss-Laguerre modes of paraxial light propagation~\cite{Siegmann-book} (where $w_0\to w_0(z)$) for $z$ values well
within the Rayleigh length $z_R=\frac{\pi}{\lambda_p}w_0^2$ of the focal plane, where $w_0(z) \approx w_0 =\text{const.}$   

We can decompose the optical polarizations $\sge(\vec r,t)$ and $\sgr(\vec r,t)$ into $u_{k,l}(r,\varphi)$ in an analogous way and obtain from 
\eqref{eq:truncated_wave_equation} and the completeness of the mode functions
\begin{multline}\label{eq:paraxial_wave_equation1}
\Bigl(\pdt+c\pdz\Bigr)\E_{kl}(z,t)=\\
ig\sum_{m,n}\int\!\mathrm dr\mathrm d\varphi\sqrt{n(\vec r)}u^*_{kl}(r,\varphi)u_{mn}(r,\varphi) \sge^{m,n}(z,t).
\end{multline}
If the atomic density $ n(\vec r) $ is
slowly varying spatially in $ r $ on the scale $ w_0 $ and furthermore is independent on $\varphi$,
orthogonality of the modes yields
\begin{equation}\label{eq:paraxial_wave_equation1a}
\Bigl(\pdt+c\pdz\Bigr)\E_{kl}(z,t)=
ig\sqrt{n} \sge^{k,l}(z,t).
\end{equation}
If furthermore the driving-field Rabi frequency  is independent on $\varphi$ and slowly varying in $r$, the
Heisenberg-Langevin equations~\eqref{eq:Heisenberg_Langevin_coherences}  decouple in the transverse modes $u_{k,l}(r,\varphi)$. In this case eq.~\eqref{eq:paraxial_wave_equation1} can be reduced to a one-dimensional propagation equation for individual transverse modes. Note that this is only correct as long as interactions are disregarded. The effect of the latter will be discussed later on. We are interested in particular in the input mode $\E_{0,0}$ for which we drop the index $(0,0)$ for simplicity in the following.

\subsection{Slow-light polaritons}

Assuming that the control field $ \Omega(z,t) $ is slowly varying in $z$ on the scale $ w_0 $  yields a system of linear partial differential equations for the probe field and the atomic coherences, the Maxwell-Bloch equations, which separate in the  transverse modes. Restricting to the lowest transverse mode and defining $\vec x=(\E,\sgr,\sge)^T$ we can write these equations as
\begin{equation}\label{eq:Hamiltonian_Matrix_realspace}
i\pdt\vec x={\mathrm H}\vec x,\quad{\mathrm H}=\begin{pmatrix}
-ic\pdz&0&-g\sqrt{n}\\
0&0&-\Omega\\
-g\sqrt{n}&-\Omega&-i\Gamma
\end{pmatrix}.
\end{equation}
Transforming to Fourier space we can find the eigenmodes of the Maxwell-Bloch equations. Restricting ourselves to small momentum $(k\approx 0)$ the eigenmodes are the so called dark-state polariton mode and two corresponding bright-state polariton modes~\cite{Fleischhauer2000}. The dark-state polariton is a superposition of electric field and atomic  coherence $\sgr$ according to the basis rotation  $\Psih=\cos\theta\E-\sin\theta\sqrt{n}\sgr$, where the mixing angle $\theta$ is given by $\tan^2\theta=g^2n/\Omega^2$. We define the bright-state polariton as $\Phih=\sin\theta\E+\cos\theta\sqrt{n}\sgr$, which is not an eigenmode of~\eqref{eq:Hamiltonian_Matrix_realspace}, but is a more convenient definition. Assuming the complex decay rate $\Gamma$ of the intermediate level to define the shortest timescale as $ |\Gamma|^{-1} $, the optical polarization $\sge$ can be adiabatically eliminated. Rotating the remaining fields $\E, \sgr$ to the basis $\vec y=(\Psih,\Phih)^T$  leads to the equation of motion
\begin{equation}
i\pdt\vec y=\mathrm H'\vec y,\
\mathrm{H'} = \mat{c\cos^2\theta\pdz &c\sin\theta\cos\theta\pdz}{c\sin\theta\cos\theta\pdz &c\sin^2\theta\pdz+\Gamma_\mathrm{eff}},
\end{equation}
where we defined the effective decay rate of the bright-state polaritons as $\Gamma_\mathrm{eff}=\Omega_e^2/\Gamma$, with $\Omega_e^2=g^2n+\Omega^2$.
Comparing the off-diagonal coupling between dark- and bright-state polaritons to the difference of the diagonal terms, we find that we can treat the off-diagonal terms perturbatively if the condition 
\begin{equation}\label{eq:EIT_window}
c|k|\ll\abs{\Gamma_\mathrm{eff}}
\end{equation}
is fulfilled for all relevant $k$-values of the polariton field. We note that this condition sets a limit for the characteristic length scale of the dark-state polaritons 
\begin{equation}\label{eq:length_dsp}
l_\mathrm{dsp}\gg\frac{c\abs\Gamma}{\Omega_e^ 2}\approx\frac{\abs\Delta}{\gamma}\La,
\end{equation}
where the second approximation holds in the off-resonant case, and $\La=g^2n/c\gamma$ denotes the {\em resonant} absorption length of the medium. It is immediately clear that many-body effects can only be observed in media with large optical depth OD$=L/L_{\rm abs}$, with $L$ being the medium length.

From the equation of motion we can read off the effective Hamiltonian in second quantization
\begin{multline}
\hat{\mathrm H}_0=-ic\int\!\!\mathrm dz\biggl[\cos^2\theta\Psid(z)\pdz\Psih(z)+\sin^2\theta\Phid(z)\pdz\Phih(z)\\
+\sin\theta\cos\theta\Bigl(\Psid(z)\pdz\Phih(z)+\Phid(z)\pdz\Psih(z)\Bigr)\\
+\frac{\Gamma_\mathrm{eff}}{c}\Phid(z)\Phih(z)\biggr],
\end{multline}
describing the free time evolution of dark- and bright-state polaritons, respectively, and their mutual coupling.

\subsection{Rydberg interactions}

For highly excited Rydberg states $\ket r$, dipole-dipole interactions between atoms become important, due to their large dipole moments~\cite{Saffman2010}. Atoms in a state $\kr$ interact with the van der Waals interaction potential $V(r)=C_6/\abs{r}^6$ with interaction strength $C_6$. For an ensemble of Rydberg atoms the microscopic Hamiltonian describing the interaction reads
\begin{equation}\label{eq:vdw_microscopic}
\hat V = \frac{1}{2}\sum_{i,j\neq i}\srr^{(i)}V(\vec r_i-\vec r_j)\srr^{(j)},
\end{equation}
where $\srr^{(i)}$ denotes the projection operator to the Rydberg state of atom $i$ at position $\vec r_i$. The interaction potential~\eqref{eq:vdw_microscopic} leads to  a space-dependent level shift which is the dominating energy scale on small length scales. If the atoms sit too close to each other the transition from a single Rydberg excitation to the doubly-excited state by laser light is prohibited, leading to a strong suppression of a second excitation, which is the so-called Rydberg blockade~\cite{Lukin2001}. 

Assuming a small excitation probability per atom to the Rydberg state, which allows to
set $\srr\approx\sgr^\dagger\sgr$, transforming to coarse-grained operators according to equation~\eqref{eq:coarse_graining} and performing the continuum limit as above 
leads to the continuous interaction Hamiltonian 
\begin{equation}\label{eq:interaction_potential_spin-operators}
\hat{\mathrm H}_\mathrm{int}=\frac{n^2}{2}\int\!\! \mathrm d^3 \vec r \!\int\!\!\mathrm d^3\vec r'\ V(\vec r\!-\!\vec r')\sgr^\dagger(\vec r)\sgr^\dagger(\vec r')\sgr(\vec r')\sgr(\vec r).\!
\end{equation}

The atom-light interaction, eqs.~\eqref{eq:Heisenberg_Langevin_coherences}, drives the atoms into 
stationary dark states, i.e., states from which there are no spontaneous emission losses, provided that
the two-photon detuning $\delta$ is sufficiently small. As a consequence the atomic medium becomes
transparent (EIT), and light propagates undepleted with reduced group velocity. The van der Waals interaction between Rydberg excitations gives rise to a two-photon level shift, which exceeds the
EIT linewidth when the distance becomes less than the EIT blockade radius
\begin{equation}\label{eq:EIT-blockade-radius}
a_B = \sqrt[6]{\abs{C_6\Gamma}/\Omega^2}.
\end{equation}
As a consequence for short distances the mixing between dark- and bright-state polaritons becomes strong and the polariton picture is no longer adequate~\cite{Bienias2014,UnanyanPrep}. However, if the excitation density is sufficiently smaller than $a_B^{-3}$ the polariton picture is expected to hold true. It should be noted that under slow-light conditions $\Omega^2$ becomes small and thus the EIT blockade radius becomes large.  In fact in the limit of light storage, where $\Omega(t)\to 0$ one finds $a_B \to \infty$ and thus it is unclear if light storage in a Rydberg EIT medium is possible at all. We will show below, however, that the critical distance $a_c$ between excitations at which a significant mixing between polaritons sets in is {\it not} given by~\eqref{eq:EIT-blockade-radius} but by
\begin{equation}\label{eq:critical_distance}
a_c = \sqrt[6]{\abs{C_6\Gamma}/\Omega_e^2},
\end{equation}
which stays finite in the limit of light storage, as $ \Omega_e^2\rightarrow g^2 n$.

\subsection{Reduction to a one-dimensional model}

Rydberg interactions can lead to a scattering between different transverse Laguerre-Gaussian modes $u_{k,l}(r,\varphi)$ and thus even the paraxial propagation of slow-light polaritons in a Rydberg gas becomes in general a three-dimensional problem. In the following we will show that also in the presence of interactions the effective one-dimensional description is valid, provided that the beam waist of the beam is small compared to the Rydberg blockade radius $a_B$. 

To derive the interaction Hamiltonian of an effective one-dimensional model we use the decomposition of the spin-flip operators $ \sgr(\vec r) $ into Laguerre-Gaussian modes 
$\sgr(\vec r)=\sum_{k,l}u_{kl}(\vec r)\sgr^{kl}(z)$ 
as in section \ref{subsec:paraxial_light_propagation}. Thus we can decompose the interaction into different potentials describing interaction between different transversal modes
\begin{multline}
\Hrm_\mathrm{int}=\frac{n^2}{2}\int\!\!\mathrm dz\!\int\!\!\mathrm dz'\!\!\sum_{\substack{k_1k_2k_3k_5\\ l_1l_2l_3l_4}}\!\!\widetilde{V}_{l_1l_2l_3l_4}^{k_1k_2k_3k_4}(z-z')\\
\times\sgr^{\dagger k_1l_1}(z)
\sgr^{\dagger k_2l_2}(z')\sgr^{k_3l_3}(z')\sgr^{k_4l_4}(z),
\end{multline}
where the effective potentials are defined by integrating out  $ r,r' $ and $ \varphi,\varphi' $,
\begin{multline}
\widetilde{V}_{l_1l_2l_3l_4}^{k_1k_2k_3k_4}(z-z'):=
C_6\int\limits_0^{2\pi}\!\!\mathrm d\varphi \int\limits_0^{2\pi}\mathrm d\varphi' \int\limits_0^\infty r~\mathrm dr \int\limits_0^\infty  r'\mathrm dr'~
\\
\times\frac{u^*_{k_1l_1}(\vec r)u^*_{k_2l_2}(\vec r')u_{k_3l_3}(\vec r')u_{k_4l_4}(\vec r)}{\Bigl[r^2+r'^2+2rr'\cos(\varphi-\varphi')+(z-z')^2\Bigr]^3}.
\end{multline}
The angular integrals can be calculated analytically by residue integration yielding 
\begin{equation}
\widetilde{V}_{l_1l_2l_3l_4}^{k_1k_2k_3k_4}(z-z')\sim\delta_{k_1-k_4,k_3-k_2} 
\end{equation}
The further evaluation has to be done numerically. We are interested in the scattering processes of an initial Gaussian mode with zero angular momentum into higher order Laguerre-Gaussian modes which are governed by the potentials $ \widetilde{V}_{l_1l_200}^{k,-k00} $. The modes $ u_{kl} $ with $ k\neq 0 $, i.e., higher-order azimuthal modes, have vanishing amplitude at $r=0$ and the corresponding interaction processes are suppressed compared to the $ k=0 $ modes. Thus we will restrict the following discussion to scattering processes into azimuthally symmetric modes. 
%
\begin{figure}[htb]
 \centering
\includegraphics[width=\columnwidth]{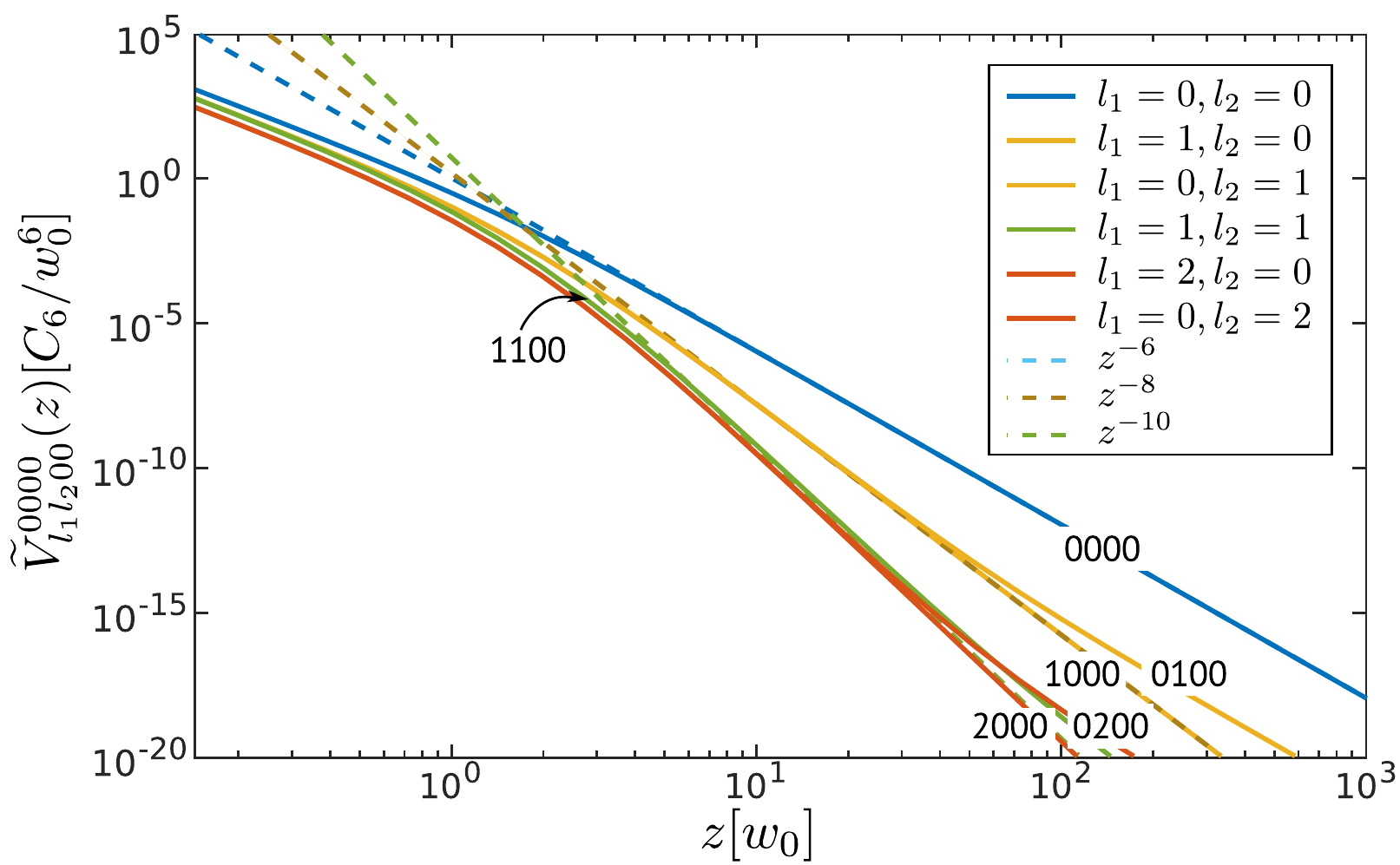}
 \caption{Different interaction potentials $ \widetilde V_{l_1l_2l_3l_4}^{0000} $ between two photons initially in Laguerre-Gaussian modes $ l_3,l_4 $ sitting at positions $(z,z'=0)$ and finally in Laguerre-Gaussian modes $l_1,l_2$, as indicated by the inset. In particular we choose initial Gaussian modes $ l_3=l_4=0 $. The dashed lines indicate power-law fits in the regime $ w_0<z<z_R $ obtained by assuming zero beam divergence $ w(z)=w_0$. }
 \label{fig:interaction_potentials} 
\end{figure}
%
In Fig.~\ref{fig:interaction_potentials} we show different potential curves for the interaction of two photons at relative distance $z,$ initially in the Gaussian mode $ l_3=l_4=0 $ and finally in (higher order) Laguerre-Gaussian modes with radial indices $ l_1, l_2 $. The relevant reference is the forward scattering potential $\widetilde V_{0,0,0,0}^{0,0,0,0}$. We find that for distances smaller than the beam waist $ w_0 $ all potentials show a $ z^{-4} $ dependence with different amplitudes due to the small overlap of the different modes. At $ z=w_0 $ the potentials cross over to a power law $z^{-\alpha}$, where we find numerically that $ \alpha\approx 6+2(l_1+l_2) $, before they flatten again for $ z>z_R $. For strong interactions small distances are blockaded, where typically the blockade distance is on the order of a few $ w_0 $. Therefore the relevant regime is given by $ w_0<a_B \le z<z_R $, where the modes functions $ u_{kl}(\vec r) $ are approximately constant in $ z$-direction and the potentials scale as $ z^{-\alpha} $. One recognizes that already for distances slightly larger than the beam waist the scattering amplitudes into higher Laguerre-Gaussian modes are orders of magnitude smaller than the forward scattering one. (Note that the potentials describing scattering between different excited modes $ u_{kl} $ are suppressed by at least an order of magnitude more and decay much faster compared to the interaction between the initial Gaussian modes.) Thus we can safely neglect the scattering processes into higher modes if the distance between excitations is sufficiently larger than the blockade radius $a_B > w_0$. This yields an effective {\it one-dimensional} interaction described by the van der Waals potential. Omitting again the indices $0$ we find
\begin{equation}\label{eq:effective-1D-interaction}
\Hrm_\mathrm{int}=\frac{n^2}{2}\int\!\mathrm dz\!\int\!\mathrm dz'~ \widetilde V(z-z')\sgr^\dagger(z) \sgr^\dagger(z')\sgr(z')\sgr(z)
\end{equation}
Note that the 1D operators $ \sgr(z)=\sgr^{00}(z) $ have a different physical dimension as the spin flip operators $ \sgr(\vec r) $ in three dimensions.

\subsection{Effective model for Rydberg dark-state polaritons}
Transforming the 1D interaction Hamiltonian~\eqref{eq:effective-1D-interaction} to the polariton basis according to $ \sqrt{n}\sgr(z)=-\sin\theta\Psih(z)+\cos\theta\Phih(z) $ yields 
\begin{multline}
\Hrm_\mathrm{int}\!=\!\frac{1}{2}\int\mathrm dz\!\int\mathrm dz'~V(z-z')\Bigl\{\bigl[\sin\theta\Psid(z)-\cos\theta\Phid(z)\bigr]\\
\times \bigl[\sin\theta\Psid(z')-\cos\theta\Phid(z')\bigr]
\bigl[\sin\theta\Psih(z')-\cos\theta\Phih(z')\bigr]\\
\times\bigl[\sin\theta\Psih(z)-\cos\theta\Phih(z)\bigr]\Bigr\}.
\end{multline}
After expanding one can identify terms describing: (i) a two-body interaction of dark-state polaritons with relative strength $ \sin^4\theta $, (ii) an interaction of bright-state polaritons with relative strength $ \cos^4\theta $ and (iii) nonlinear coupling terms of dark- and bright-state polaritons. Note that in the slow light regime $ 1\approx\sin^4\theta\gg\cos^4\theta $ the dark-state polaritons, consisting mainly of Rydberg excitation, are strongly interacting while the bright-state polaritons, consisting mainly of electric field excitation, are weakly interacting. In lowest order in $\cos\theta$ only a nonlinear interaction term between Rydberg dark-state polaritons  remains. In the following we want to consider also the dominant correction terms, caused by the coupling of the dark-state polaritons to bright-state polaritons, which we can assume to be in 
a vacuum state.

The full time evolution of dark- and bright-state polaritons is governed by the sum of free and interaction Hamiltonian and can analogously be decomposed into dark-state, bright-state and coupling terms, $\Hrm=\Hrm_0+\Hrm_\mathrm{int}=\Hrm_{\Psi}+\Hrm_{\Phi}+\Hrm_{\Psi\Phi}$. We make a standard system-plus-reservoir approach~\cite{CarmichaelBook} and derive an effective Liouvillian for the Rydberg dark-state polaritons by treating the bright-state polaritons as a Markovian reservoir in the vacuum steady state. The free time evolution of the polaritons in zeroth order in the coupling is described by the differential equations
\begin{align*}
\pdt\Psih(z,t)&=-c\cos^2\theta\pdz\Psih(z,t)\\
\pdt\Phih(z,t)&=-c\sin^2\theta\pdz\Phih(z,t)-\Gamma_\mathrm{eff}\Phih(z,t),
\end{align*}
where we have disregarded the Langevin noise terms.
As can easily be derived, the solution of these equations is given by
$\Psih(z,t)=\Psih(z-c\cos^2\theta\tau,t-\tau)$ and 
$\Phih(z,t)=e^{-\Gamma_\mathrm{eff}\tau}\Phih(z-c\sin^2\theta\tau,t-\tau)$.
Thus the free dark-state polaritons propagate with group velocity $ \vg=c\cos^2\theta $ with a stable shape, while the bright-state polaritons propagate with $ c\sin^2\theta $ and are subject to decay with the rate  $\mathrm{Re}[\Gamma_\mathrm{eff}]$. Assuming that 
$ |\Gamma_\mathrm{eff}|^{-1} $ defines the fastest time scale, justifies adiabatic elimination of the bright-state polaritons. If there is no external driving of bright-state polaritons, the steady state is the vacuum and we find that bright-state polaritons are delta-correlated in space and their correlation functions decay in time as
\begin{align}
&\braket{\Phih(x,t)\Phid(y,t-\tau)}\approx  e^{-\Gamma_\mathrm{eff}\tau}\delta(x-y),\quad\ \tau>0,\\
&\begin{gathered}[b]
\braket{\Phih(x,t)\Phih(x',t)\Phid(y,t-\tau)\Phid(y',t-\tau)}\approx \\ 
\approx e^{-2\Gamma_\mathrm{eff}\tau} \Bigl[\delta(x-y-v\tau)\delta(x'-y'-v\tau)\\
\qquad\quad+\delta(x'-y-v\tau)\delta(x-y'-v\tau)\Bigr],
\end{gathered}\tau>0.
\end{align}
All remaining correlation functions of bright-polariton operators vanish in the vacuum state. 

In an interaction picture with respect to the system and reservoir Hamiltonian $\Hrm_{\Psi}+\Hrm_{\Phi}$ the dark-state polariton degrees of freedom are described by the density matrix
 $ \rho=\mathrm{tr}_\Phi\chi$, which we get by partial trace of the full density matrix $ \chi $ over the bright-state polaritons. The time evolution of $ \rho $ is then governed in Born approximation by the equation
\begin{equation}\label{eq:born_markov}
\dot\rho_\Psi(t)=-\int_0^\infty\mathrm\!\!\!\! d\tau\ \mathrm{tr}_\Phi\left\{\Bigl[\Hrm_{\Psi\Phi}(t),\bigl[\Hrm_{\Psi\Phi}(t),\rho(t)\otimes\rho_\Phi\bigr]\Bigr]\right\},
\end{equation}
where $\rho_\Phi$ denotes the bright-state polariton steady state density matrix. 
We note that the free part of $\Hrm_{\Psi}$ corresponds to a transformation to a frame comoving with the group velocity $ \vg=c\cos^2\theta $. Introducing the operator
\begin{equation}\label{eq:Lindblad_generator}
\hat L\equiv-\sin^3\theta\cos\theta\Biggl[\int\! \mathrm ds V(z-s)\Psid(s)\Psih(s)
+i\frac{c\pdz}{\sin^2\theta}\Biggr]\Psih(z),
\end{equation}
allows us to write the system-reservoir-coupling Hamiltonian in the interaction picture in the compact form
\begin{multline}\label{eq:H_bsp_dsp_coupling_L}
\Hrm_{\Psi\Phi}(t)=\intdz\ \Bigl\{ \Phid(z)\hat L(z)	+\hat L^\dagger\Phih(z)
+\intdz'V(z-z')\\
\times\sin^2\theta\cos^2\theta
\left( \hat\Phi^\dagger(z,t)\hat\Phi^\dagger(z',t)\Psih(z',t)\Psih(z,t)\right.\\
 + \left.\Psid(z,t)\Psid(z',t)\hat\Phi(z',t)\hat\Phi(z,t) \right)\Bigr\},
\end{multline}
where we omitted terms third order in the bright-state polaritons since the three particle correlations of bright-state polaritons are vanishing in the vacuum.
We insert Hamiltonian~\eqref{eq:H_bsp_dsp_coupling_L} into equation~\eqref{eq:born_markov} and follow the standard derivation of a Master equation~\cite{CarmichaelBook}. By performing the Markov approximation we get a Master equation in Lindblad form describing an effective time evolution of the dark-state polaritons. After transforming back to the moving frame Schr\"odinger picture we arrive at
\begin{widetext}
\begin{multline}\label{eq:Lindblad_Master_equation}
\dot\rho=i\frac{\Delta}{\Omega_e^2}\intdz\left[\rho,\hat L^\dagger(z)\hat L(z)\right]
+i\int\mathrm dz\int\mathrm dz'~V(z-z')\left[\rho,\Psid(z)\Psid(z^\prime)\Psi(z^\prime)\Psi(z)\right]
\\
+i\frac{\Delta\sin^4\theta\cos^4\theta}{\Omega_e^2}\intdz\intdz'\ V^2(z-z')\left[\rho,\Psid(z)\Psid(z')\Psih(z')\Psih(z)\right]
+\frac{\gamma}{\Omega_e^2}\intdz\left(2\hat L(z)\rho\hat L^\dagger(z)-\left\{\rho,\hat L^\dagger(z)\hat L(z)\right\}\right)
\\
+\frac{\gamma\sin^4\theta\cos^4\theta}{\Omega_e^2}\intdz\intdz'\ V^2(z-z')
\left(2\Psih(z')\Psih(z)\rho\Psid(z)\Psid(z')-\left\{\rho,\Psid(z)\Psid(z')\Psih(z')\Psih(z)\right\}\right)
\end{multline}
\end{widetext}
where $\{\cdot,\cdot\}$ denotes the anti-commutator. 
The coupling of dark- and bright-state polaritons gives rise to various effective unitary and dissipative terms in equation ~\eqref{eq:Lindblad_Master_equation} influencing the evolution of dark-state polaritons.
Specifically, we find the following unitary terms: a kinetic energy term with an effective mass defined by
\begin{equation}\label{eq:effective_mass}
m^{-1}=2\vg\frac{c\Delta\sin^2\theta}{\Omega_e^2}=2\vg\frac{\Delta}{\gamma}\La\sin^4\theta,
\end{equation}
a drift term $\sim\intdz'V(z-z')\Psid(z')\Psih(z')$ -- mediated by interactions -- giving corrections to the group velocity $\vg$, and, finally,  higher order terms in the interaction, namely a three body interaction and corrections to the two-body interaction $\sim V^2(z-z')$. 

Moreover, since the bright-state polaritons decay the coupling also leads to effective loss channels for the dark-state polaritons in equation~\eqref{eq:Lindblad_Master_equation}. The decay processes are described by the operator $ \hat L $ defined in~\eqref{eq:Lindblad_generator}. This operator consists of two terms arising from the two effective loss channels describing coupling between dark- and bright-state polaritons, namely a linear coupling outside the EIT window and a nonlinear coupling arising from the interaction. Both terms lead to losses of dark-state polaritons. In particular, we identify a generalized single particle loss generated by the Lindblad operator $\sqrt{\hat\Gamma(z)}\Psih(z)$ with the operator valued loss rate
\begin{equation}\label{eq:operator_loss_rate}
\hat{\Gamma}(z)=\frac{\gamma\cos^2\theta\sin^6\theta}{\Omega_e^2}\left[\int\! \mathrm dx\ V(z-x)\Psid(x)\Psih(x)\right]^2\!\!.
\end{equation}
Note that this loss as well as the corrections to the drift term cannot be simply treated in a mean-field approximation, since $V(z-z')$ can assume arbitrary values. However, the full Lindblad operator gives rise to a space-dependent loss process $\sim r^{-12}$ that is strong for small distances $r$. 



\subsection{Effective Hamiltonian}

After an initial transient we expect polariton states to evolve according to equation~\eqref{eq:Lindblad_Master_equation} and to show an blockaded volume for small distances. We assume that the product of interaction energy and mass is always positive, corresponding to repulsive interaction that prevents excitations to propagate into the blockade volume. On large length scales the interaction decays as $ V(r)\sim\abs{r}^{-6} $, thus the effective interaction terms linear in $ V $ are the dominant contributions and we may neglect the two body interaction with potential $ \sim V^2\sim \abs{r}^{-12} $ and the three-body interaction as well as the loss term with rate~\eqref{eq:operator_loss_rate}. As a second approximation we assume that we are in the slow-light regime, where  $ \cos\theta\ll 1 $ and only excitation energies inside the EIT window are allowed. Finally, we consider an off-resonant driving scheme where the fields are far detuned and $ \gamma\ll\Delta $ which allows us to neglect the remaining loss terms. An expansion in $ \cos\theta $ up to second order reduces the master equation to a von Neumann equation for the  density matrix, governed by the Hamiltonian
\begin{multline}\label{eq:H_polariton_interacting}
\Hrm=-\intdz\ \Psid(z)\frac{\pdz^2}{2m}\Psi(z)\\
+C_6\sin^4\theta\intdz\intdz^\prime\frac{\Psid(z)\Psid(z^\prime)\Psi(z^\prime)\Psi(z)}{\abs{z-z^\prime}^6},
\end{multline}
in the moving frame Schr\"odinger picture.

\section{Two-particle simulations and verification of effective model}
\label{sec:two-particle-simulations}

\begin{figure*}
 \centering
\includegraphics[width=2\columnwidth]{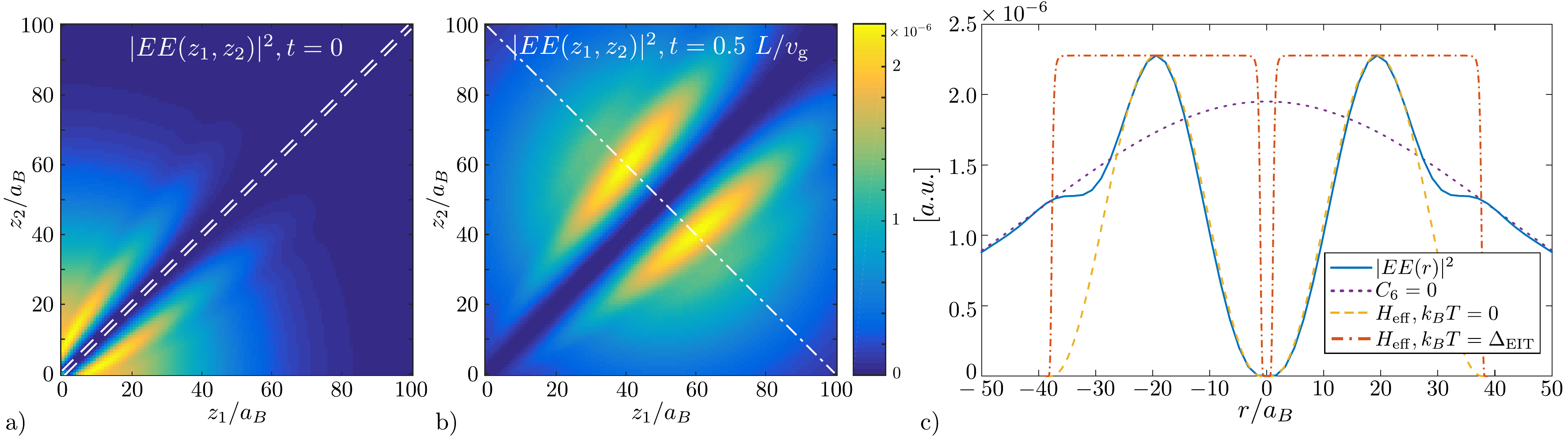}
 \caption{
 Splitting of initial Gaussian wave packet due to van der Waals interaction. Full two-excitation wave function simulation for the parameters $g=10\Omega=10\gamma$ and $\Delta=4\gamma$. The optical depth per blockade is chosen $\ODB \equiv a_{\mathrm B}/L_{\rm abs}= 7$. The color plots in a) and b) show the two-excitation probe field amplitude $EE(z_1,z_2,t)$ at times $t=0$ and $t=0.5L/\vg$ corresponding to the times where the center of the incident Gaussian wave packet is at the medium boundary and at the middle of the medium, resp. Plot c) shows a cross section (blue solid line) of the middle color plot along the diagonal $r=z_1\!-\!z_2$ for $2R =z_1\!+\!z_2=\!L\!=\! 100~a_{\mathrm B}$, as indicated by the dashed-dotted line in b).  We compare the cross section to a pulse propagating in absence of interactions (Gaussian, purple dotted line) and the two-particle ground state of the effective Hamiltonian~\eqref{eq:H_polariton_interacting} (yellow dashed line),where we impose periodic boundary conditions with a period corresponding to the distance of the two peaks (blue solid line). The red dash-dotted line shows a thermal two-particle state of the effective Hamiltonian. For illustration purposes we continued the two-particle ground state and the thermal state.}
 \label{fig:initial_state}
\end{figure*}

We are interested in the effect of the Rydberg interactions on the evolution of propagating dark-state polaritons. For two copropagating polaritons this has been considered theoretically and simulated numerically in~\cite{Gorshkov2011} and, in a resonant as well as  off-resonant regime recently been demonstrated in experiments~\cite{Peyronel2012,Firstenberg2013}. We want to extend this to the strongly interacting far off-resonant regime, where we predicted in the previous section that the dynamics is governed by the effective Hamiltonian ~\eqref{eq:H_polariton_interacting} in a comoving frame.

It was shown in~\cite{Gorshkov2011} that the van der Waals interaction leads to an avoided volume for the propagating photons of  size $2a_B$. Near single-photon resonance, $|\Delta| \leq \gamma$, two photons at distance smaller than $a_B$ get absorbed and an incident two-photon wave-packet evolves into a non-classical state that is a statistical mixture of a single excitation and a correlated train of (two) photons separated by $2a_B$~\cite{Gorshkov2011}. In the off-resonant regime ($\abs\Delta\gg\gamma$) the absorption plays no role but two photons closer than the
blockade distance propagate with the vacuum speed of light and thus escape. Moreover, the repulsive interaction prevents the photons to get inside the blockade radius.

To validate the simple picture we perform numerical simulations of the full 1D Maxwell-Bloch equations for two copropagating polaritons subject to van der Waals interaction.  The two-excitation wave function $\ket{\Psi_2(t)}$ is composed of six two-excitation modes describing all combinations of the single-excitation modes $\E,\sgr$ and $\sge$.  
Note that in presence of decay the norm of the wave-function is not conserved. 

 We consider an initial Gaussian two-photon wave packet $\bigl(\ket{\Psi_0}=\Ed\Ed\ket{0}\bigr)$ in free space, where we assume that the spectral width fits well inside the EIT window of the medium in absence of interactions. The state vector at a later time $t$ is then described by two-excitation wave functions
 \begin{eqnarray}
 \ket{\Psi(t)} = \int\!\! {\rm d}z_1 \int\!\! {\rm d}z_2 \sum_{F,G} FG (z_1,z_2,t) \hat F^\dagger(z_1) \hat G^\dagger(z_2)
 |0\rangle\nonumber
 \end{eqnarray}
 where $FG\in \{EE,PP,SS,EP,PE,ES,SE,PS,SP\}$ denote all possible two-particle wave functions corresponding to excitations
 in the electric field $(E)$, the optical polarization $(P)$, and the spin polarization $(S)$.
 We then simulate the propagation of the pulse from free space into the Rydberg gas by considering a space-dependent atomic coupling strength $g^2n=g^2n(z)$, where we use a step function for $ n(z) $. To take care of the EIT pulse compression for small group velocities we use an adaptive spatial grid spacing.
In Fig.~\mbox{\ref{fig:initial_state} a), b)} we show two snapshots of the time evolution of the $E\!E(z_1,z_2)$-component incident into the medium at the lower left corner ($z_{1/2}=0$) and propagating along the diagonal to the upper right corner of the pictures. As can be seen from the figure in the considered off-resonant regime the wave packet splits at the medium boundary into two parts separated by the off-resonant blockade radius $a_B$, indicating a photon antibunching as shown for the resonant case in~\cite{Gorshkov2011}. However, after longer propagation inside the medium the separated parts move even further apart due to the repulsive interaction on length scales $\Delta z\gg a_B$.

The results in Fig.~\ref{fig:initial_state} verify the expected time evolution governed by Hamiltonian~\eqref{eq:H_polariton_interacting}. This can be seen from Fig.~\mbox{\ref{fig:initial_state} c)}, where we compare a cross section (blue solid line) of the evolved pulse  to the two-excitation ground state of the effective Hamiltonian (yellow dashed line) calculated by employing periodic boundary conditions. For illustrative purposes we have periodically continued the ground-state wave-function in the plot (yellow dash-dotted line).

We recognize two important things: First, we find a remarkably good agreement of the numerically evolved wave-packet to the ground state of the effective Hamiltonian up to a length scale of $ \sim\pm 25a_B $. Secondly, contrary to~\cite{Firstenberg2013}, we do not find a localized two-particle excitation around $r=0$ (bound state). That the final state of the evolution is so close to the ground state is somewhat unexpected since the initial wave-packet contains many highly excited components. The presence of a finite EIT frequency window sets some upper bounds on the energy of the polaritons. Since the system is not integrable, we can assume that the polaritons quickly thermalize after entering the interacting medium. However, as can be seen from comparison with the red (dash-) dotted curves in Fig.~\ref{fig:initial_state}, estimating the temperature by the EIT transmission window~\eqref{eq:EIT_window} is much too high. A possible explanation of this is the presence of a mechanism similar to evaporative cooling. The high-energy modes of the Rydberg dark-polariton modes are preferentially coupled to bright-polariton modes and decay or propagate away. A full analysis of this mechanism is however not subject of the present paper.

Let us briefly comment on the fact that we do not see bunching of photons in the numerical simulations of the full Maxwell-Bloch equations
as in~\cite{Firstenberg2013}. We here consider the case $mC_6>0$, where the effective Hamiltonian \eqref{eq:H_polariton_interacting} does not
have bound states which could explain the observed bunching. The full set of Maxwell-Bloch equations on the other hand should reproduce the bunching.
However, our simulation is run in a regime with much larger optical depth per blockade $\ODB$ as in~\cite{Firstenberg2013}. As will be discussed elsewhere\cite{UnanyanPrep}, in this regime not a single, but many bound states exist which cannot be excited, however, by the incident photon pulse. For values of $\ODB$ much smaller than considered in the present paper we also find bunching in our numerics.

\section{Ground state properties of effective Hamiltonian and Luttinger liquid Approach}

We have seen in the previous section that a finite-length light pulse entering an EIT medium with Rydberg interactions with large optical depth per blockade $\ODB$ generates a many-body state of Rydberg dark-state polaritons with an energy close to the ground state of the effective Hamiltonian~\eqref{eq:H_polariton_interacting} in the moving frame. In this section we will discuss the ground-state properties of this Hamiltonian using numerical DMRG simulations as well as a Luttinger-liquid approach.

\subsection{DMRG ground state simulations}\label{subsec:DMRG}

\begin{figure}
 \centering
\includegraphics[width=\columnwidth]{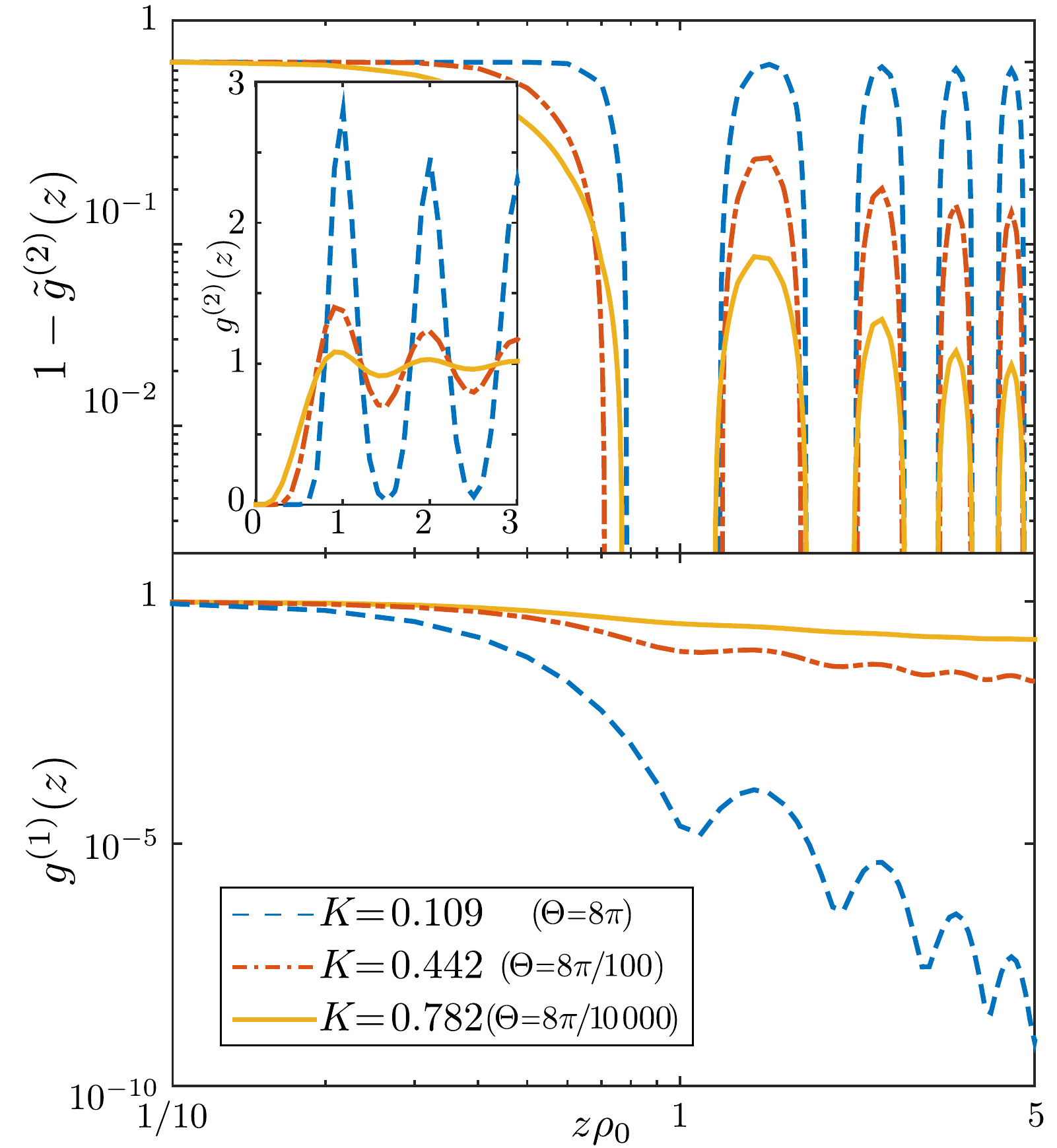}
 \caption{DMRG correlation functions. \textit{Top:\ }1-density-density correlation functions for different interaction strength in a double logarithmic plot and density-density correlation functions in a linear plot in the inset. Taken from~\cite{Otterbach2013}. \textit{Bottom:\ }corresponding first order correlation functions.}
 \label{fig:DMRG_correlations}
\end{figure}

The physics of~\eqref{eq:H_polariton_interacting} is governed by an interplay of kinetic energy contribution trying to delocalize the massive particles and polariton-polariton interaction leading to spatial order. Note that the effective mass~\eqref{eq:effective_mass} can be negative, depending on the sign of the detuning $\Delta$. We assume in the following that the product of $m C_6$ has always a positive sign, corresponding to a repulsive interaction. To analyze the interplay between kinetic and interaction energies we calculated ground state correlations using the density matrix renormalization group~\cite{Schollwoeck2011}. Specifically we calculate single- and two-body correlation functions for different interaction strengths, quantified by the dimensionless parameter $ \Theta $, which is defined as the ratio of interaction energy at average inter-particle distance $ 1/\rho_0 $ compared to the Fermi energy~\footnote{Note1},
\begin{equation}\label{eq:Theta}
\Theta = \frac{(\rho_0a_B)^4}{4\pi}\left(\frac{\gamma}{\Delta}\right)^2\ODB^2.
\end{equation}
In Fig.~\ref{fig:DMRG_correlations} we show density-density and first order correlations calculated by DMRG. For short distances the density-density correlations are strongly suppressed, i.e., $ g^{(2)}(0)=0$, indicating the photon blockade of two copropagating excitations. For small $ \Theta $ corresponding to weak interactions the first order correlation functions govern the long-range behavior, indicating a superfluid state. Increasing the interaction strength leads to a strongly pronounced and slowly decaying density wave (CDW) with fast decaying first order correlations. From the plot we can extract a power law decay $ z^{-\beta}$ for distances larger $ 1/\rho_0 $. Note that for $z\rho_0\gtrsim 4$ finite size effects influence the results. Due to the low dimensionality ($d=1$) of the model no slower-than-power law decay of correlations can be found, i.e., the model cannot exhibit true crystalline order. However, the latter can in principle be created by engineering an additional lattice potential, i.e., a space-dependent two-photon detuning for the polaritons leading to a sine-Gordon like model for commensurate fillings. This model exhibits a quantum phase transition to a gapped phase with true crystalline order~\cite{Giamarchi-book,Buechler2011}.

\subsection{Luttinger liquid approach}

The low energy physics of a gapless one-dimensional system can be described in terms of a Luttinger liquid (LL) model. Assuming that we have a state of $N$  polariton excitations in a frame moving with group velocity $ \vg $ distributed over length $ L $ yields an excitation density of $ \rho_0=N/L $.
Working in the far off-resonant regime, spontaneous decay can be neglected after an initial transient and the repulsive interaction suppresses the coupling to the bright-state polariton loss channels. Therefore the number of polaritons can be assumed to be constant. Following standard LL theory~\cite{Giamarchi-book} we express the bosonic polariton field $ \Psih(z) $ by fields $ \phi(z) $ and $ \theta(z) $ representing density and phase fluctuations of the long-lived sound-modes, respectively. The density of the polariton field transforms as $ \hat{\rho}(z)=\Psid(z)\Psih(z)=-\pdz\phi(z)/\pi$. Writing the system Hamiltonian in terms of $ \phi(z) $ and the canonically conjugate mode $ \Pi(z)=\pdz\theta(z)/\pi $ yields the LL Hamiltonian 
\begin{equation}\label{eq:H_Luttinger_A}
H_\mathrm{LL}=\frac{1}{2\pi}\intdz\left[v_sK\bigl(\pi\Pi(z)\bigr)^2+\frac{u}{K}\bigl(\nabla\phi(z)\bigr)^2\right].
\end{equation}
The Hamiltonian is fully defined by two parameters, the speed of sound $v_s$ and the Luttinger parameter $K$. The $K$ parameter universally determines the long-range behavior of the ground state correlation functions. In particular, the equal time density-density correlations are given by~\cite{Giamarchi-book}
\begin{equation}\label{eq:density_correlations_luttinger}
\braket{\rho(z)\rho(0)}=\rho_0^2-\frac{K}{2\pi^2}\frac{1}{z^2}+A_2\rho_0^2\cos(2\pi\rho_0z) e^{-2G_{\phi\phi}(z)},
\end{equation}
where $G_{\phi\phi}(z)\equiv \braket{[\phi(z)-\phi(0)]^2}$ and $A_2$ is a non-universal amplitude that has to be determined numerically from the microscopic model. We are particularly interested in the last term that oscillates spatially with the period $1/\rho_0$ corresponding to a charge-density wave (CDW). The spatial decay of the CDW is governed by the correlation function $e^{-2G_{\phi\phi}(z)}$ which in the ground state is given by a power law $(\alpha/z)^{2K}$ with exponent $2K$ and a short distance cutoff $\alpha$, which we choose to be the shortest length scale which is typically the average distance of excitations $\rho_0^{-1}$. In contrast to this the ground state first order correlation function describing superfluid order is given by
\begin{equation}\label{eq:fo_correlations_luttinger}
\braket{\Psid(z)\Psi(0)}=\rho_0A_1\left(\frac{\alpha}{z}\right)^{1/2K},
\end{equation}
where the amplitude $A_1$ is also non-universal. 

\subsection{Quasi-Crystalline state under stationary EIT conditions}

The long-range behavior of first order~\eqref{eq:fo_correlations_luttinger} and density-density correlations~\eqref{eq:density_correlations_luttinger} are in the ground state both given by a power law with exponents $1/2K$ and $2K$, respectively. Comparing this with Fig.~\ref{fig:DMRG_correlations} shows that they both agree qualitatively well with the DMRG results.  Depending on $K$ either superfluid (first-order) ($K\gg 1/2$) correlations or the CDW ($K\ll 1/2$) dominate the long-range nature of correlations, where the point $K=1/2$ marks the crossover point, where both correlations decay with an exponent $1$. For realization of a quasi-crystalline state a regime $ K\ll 1 $ has to be realized.
Notice that the long-range interaction yields a $K$-parameter smaller $1$ for a sufficient interaction strength, indicating more pronounced density correlations than free fermions which is the strongest that can be achieved for $\delta$-interacting bosons~\cite{Giamarchi-book}, e.g., photons interacting via a Kerr nonlinearity~\cite{Chang2008,Angelakis2011}.

For our model~\eqref{eq:H_polariton_interacting} the dependence of the $K$-parameter as a function of microscopic parameters cannot be given analytically, but has to be determined numerically. We can extract $K$ from DMRG simulations, where we calculated the compressibility $\chi^{-1}=\rho_0^2\frac{\partial\mu}{\partial\rho_0}=\rho_0^2L\frac{\partial^2E}{\partial N^2}$. From this we get the ratio $K/v_s=\pi\rho_0^2\chi$ and together with the product $v_sK=\pi\rho_0/m$ which is constant for any Galilean invariant system~\cite{Giamarchi-book} we can extract the $K$-parameter as a function of the dimensionless parameter $\Theta$,
\eqref{eq:Theta}. In the case of power law interactions an approximate analytical formula for the $K$-parameter has been given in~\cite{Dalmonte2010}, which is asymptotically correct for small and large $\Theta$, 
\begin{equation}\label{eq:K_Parameter_ODB}
K = \left(1+\tfrac{\pi^4}{45}\Theta\right)^{-1/2}.
\end{equation}
%

\begin{figure}
	\centering
	\includegraphics[width=\columnwidth]{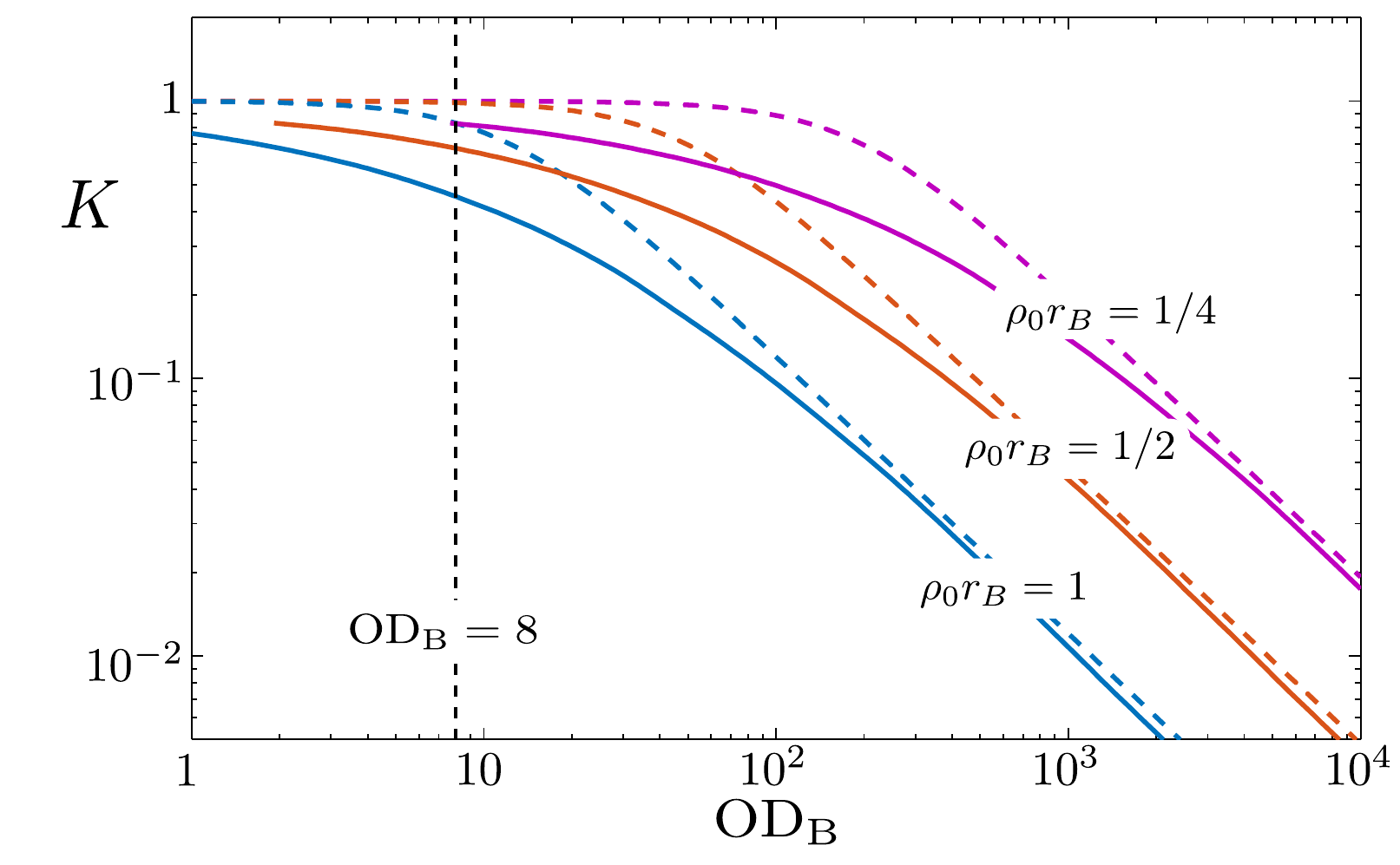}
	\caption{%
		$K$-parameter as a function of the optical depth per blockade $\ODB$ for $\Delta/\gamma=10$. Solid curves are interpolated numerical data from DMRG results, corresponding dashed curves are approximate values according to~\eqref{eq:K_Parameter_ODB}. The three different colored pairs of curves are for different values $\rho_0a_B$ as indicated in the plot. The dashed vertical line indicates experimental values from~\cite{Peyronel2012}.
	}%
	\label{fig:K_ODB}
\end{figure}
From this we can estimate the experimental requirements for reaching the strongly interacting regime ($K\ll 1$) under stationary EIT conditions. An important restriction of {\it stationary} EIT is set by the Rydberg blockade: When two excitations get closer than the EIT blockade radius $a_B$,~eq.\eqref{eq:EIT-blockade-radius}, they are either absorbed or transformed into fast propagating
bright-state polaritons which escape. Thus the excitation density is limited due to photon blockade to values 
\begin{equation}
\rho_0 a_B\leq 1
\end{equation}
and in the regime of small $K$ we have $K\sim\ODB^{-1}$ indicating that even for a $\rho_0 a_B\approx1$ the required optical depth is orders of magnitude higher than experimentally feasible~\cite{Peyronel2012}, cf. Fig.~\ref{fig:K_ODB}. Moreover, increasing $\ODB$ by changing the blockade distance $a_B$ requires a smaller excitation density $\rho_0$ for fixed $\rho_0a_B$ and thus limits the number of excitations in a medium with a finite length such that in the limit $\ODB\gg 1$ only a single excitations would be allowed inside the medium for {\it stationary} EIT driving.

\section{Reaching the strongly interacting regime: light storage in Rydberg gases}

\subsection{frequency pulling in adiabatic slow-down of Rydberg polaritons}

The problems to reach the required conditions for quasi-crystalline order of Rydberg polaritons under stationary EIT conditions can be overcome using a {\it dynamical} protocol, i.e., by considering light storage or dynamical slow down of polaritons while inside the medium.  One recognizes from~\eqref{eq:EIT-blockade-radius} that during slow down and ultimately light storage the EIT blockade radius $a_B\sim \Omega^{-1/3}$ would diverge. Naively one would expect that as a consequence the smallest possible distance between Rydberg polaritons diverges as well, when the group velocity goes to zero. Remarkably this is not the case. To see this let us consider the dynamics of a dark-state polariton during storage with a finite two-photon detuning $\delta_0$, resulting, e.g., from a second
nearby Rydberg excitation. As has been shown in~\cite{Mewes2002} a small two-photon detuning causes a time-dependent phase shift (chirp) of the dark-state polariton during slow down:
\begin{multline}
\Psi(z,t) = \Psi\left(z-c\int_0^t\!\! {\mathrm d}\tau \cos^2(\tau),0\right)\\
\times\mathrm{exp}\left\{i \delta_0 \int_0^t\!\! {\mathrm d}\tau \sin^2(\tau)\right\}.
\end{multline}
As a consequence the spectrum of the probe field ${\cal E}(z,t)=\cos\theta(t)\Psi(z,t)$ assuming a slowly varying mixing angle $\theta(t)$ can be expressed as
\begin{multline}
S(z,\omega) =\int\limits_{-\infty}^{\infty} {\rm d}\tau e^{-i\omega \tau} \Bigl\langle {\cal E}^\dagger(z,t){\cal E}(z,t-\tau)\Bigr\rangle \\
= S\left(0,\frac{1}{\cos^2\theta(t)}\left(\omega -\delta_0 \sin^2\theta(t)\right)\right).
\end{multline}
One recognizes two effects: First, there is a spectral narrowing proportional to $\cos^2\theta(t)$, which guarantees that during light storage the pulse spectral width remains less than the EIT transparency window~\eqref{eq:EIT_window}, if it did so at the beginning of the storage process~\cite{Fleischhauer2002}. Secondly, and more importantly in the present context, there is a pulling of the center frequency of the pulse towards the two-photon resonance
\begin{equation}
\delta(t) = \delta_0 \cos^2\theta(t).
\end{equation}
This effect has been observed experimentally in~\cite{Karpa2009} and is responsible for the fact that the two-photon linewidth of EIT light storage~\cite{Mewes2002}
\begin{equation}
\delta_{\mathrm 2ph} = {\Omega_e^2}/{|\Gamma|}
\end{equation}
is determined by the collective Rabi frequency $\Omega_e$ rather than the control-field Rabi frequency $\Omega(t)$. (Here $\Gamma = \gamma +i\Delta$ and we have set the optical depth $\OD=1$.)
\begin{figure}
	\centering
	\includegraphics[width=\columnwidth]{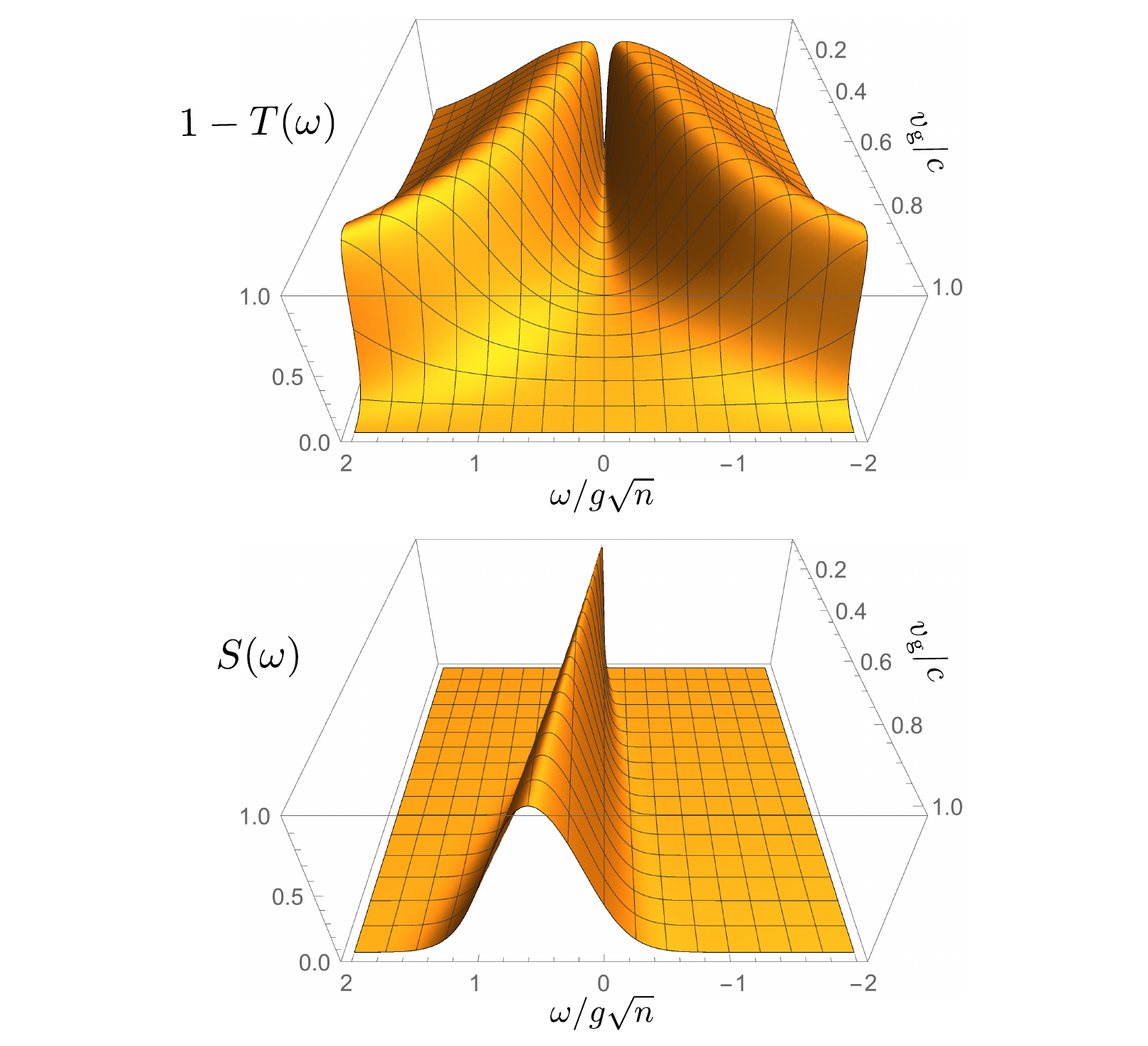}
	\caption{%
	Illustration of transmission spectrum $T(\omega)$ of the EIT medium ({\it top}) and spectrum of	the electromagnetic field ({\it bottom}) in arbitrary units  as function of the EIT mixing angle $\theta$ during a dynamical light storage protocol.
	}%
	\label{fig:pulling}
\end{figure}

\begin{figure}
	\centering
	\includegraphics[width=\columnwidth]{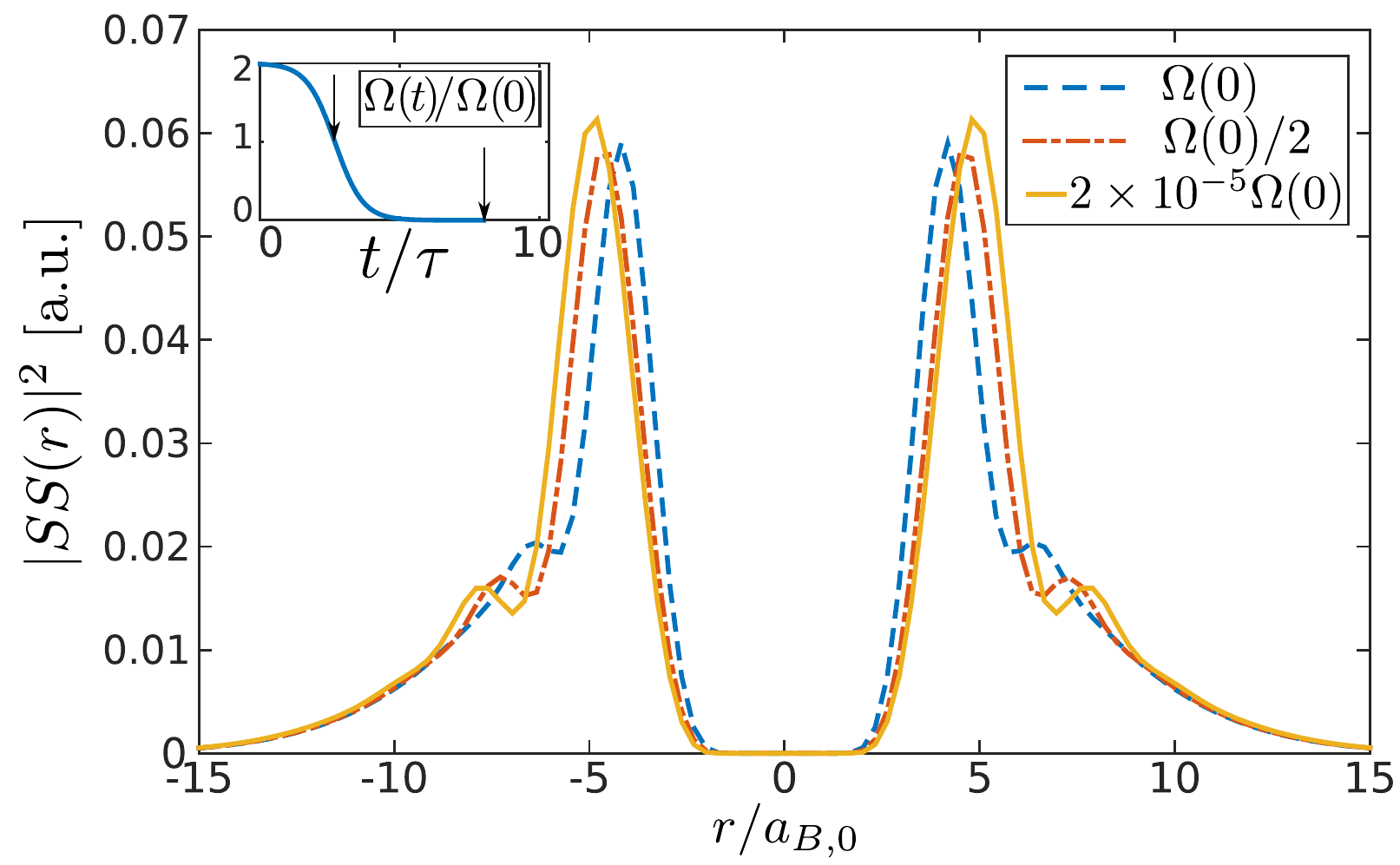}
	\caption{%
Cross section of $SS$-component during storage. The control field is turned off according to a protocol $\tanh(t/\tau)$, which is shown in the inset. The main figure shows the cross section of the doubly excited spin component for 3 different times, before starting the storage protocol, at time $t = 2\tau$ , where $\Omega(t) = \frac{1}{2}\Omega_0$ and at the end of the protocol, where $\Omega(t)$ is approximately zero.}%
	\label{fig:two-photon-light-storage}
\end{figure}

As a consequence of this the minimum distance of two Rydberg polaritons is not given by the EIT blockade distance,~\eqref{eq:EIT-blockade-radius}, but by the critical distance
\begin{equation}
a_c=\sqrt[6]{\abs{C_6\Gamma}/{\Omega_e^2}}.
\end{equation}
We have verified this by two-particle simulations, which are illustrated in Fig.~\ref{fig:two-photon-light-storage}. Shown is the spin-spin component of an initial two-photon wave-packet propagating in an EIT medium with Rydberg interactions as in section~\ref{sec:two-particle-simulations}. After an initial transient an avoided volume is formed corresponding to the blockade radius $a_B(\Omega_0)$ given by the \textit{initial~}drive-field Rabi-frequency $\Omega_0$ and larger than the critical distance
$a_c$. When the drive field is adiabatically turned to zero as shown in the inset, the EIT blockade radius $a_B(\Omega(t))$ increases and eventually diverges. The avoided volume stays however approximately constant.

Therefore adiabatic slow down of Rydberg polaritons allows to increase their effective mass and thus the ratio of interaction energy to kinetic energy, $\Theta$, without reducing the polariton density $\rho_0$. In this way it is possible to enter the interesting regime of strong interactions between Rydberg dark-state polaritons!

\subsection{Time-dependent Luttinger liquid approach}

In the previous section we have shown that by employing a time-dependent protocol a regime of strongly interacting polaritons with $K\ll 1$ can be reached, as opposed to a continuous driving. Through two-photon simulations we have validated that during storage the polaritons are separated by the finite distance $a_c$.
The ground state in the regime $K\ll 1$ is strongly correlated as we have shown in section \ref{subsec:DMRG}. However, since the Luttinger liquid is gapless the quench from weakly to strongly interacting regime cannot be fully adiabatic. Nevertheless, as will be shown in the following a state with long-range CDW correlations that extend over a finite length can be prepared.

To analyze the many-body polariton dynamics we calculate the time evolution of the correlation functions~\eqref{eq:density_correlations_luttinger} starting from a given (weakly interacting) initial state which we assume to have been prepared under stationary EIT conditions. To get the correlation functions after storage we apply a time-dependent Luttinger theory as in~\cite{Dora2011}. 

Storing the polaritons by tuning the control field $\Omega$ to zero leads  to time-dependent parameters $K(t)$ and $v_s(t)$, where we can calculate their time-dependence from the effective mass analytically using~\eqref{eq:K_Parameter_ODB} and also numerically from exact DMRG simulations (Fig.~\ref{fig:K_ODB}).
We follow standard bosonization procedure and decompose the canonically conjugate fields $\Phi$ and $\Pi$ into bosonic momentum modes $[b_q,\bd_p]=\delta_{p,q}$. This transforms the Hamiltonian~\eqref{eq:H_Luttinger_A} to
\begin{equation}\label{eq:H_Luttinger_B}
\Hrm_\mathrm{LL}=\frac{v_s(t)}{2}\sum_{p\neq 0 }\abs{p}\left[w(t)\bd_p b_p-\frac{g(t)}{2}\left(\bd_p\bd_{-p}+b_{-p}b_p\right)\right],
\end{equation}
where $w(t)$ and $g(t)$ are given by the LL parameter as  $K(t)\pm K^{-1}(t)$, respectively. The time evolution of the system can then be calculated by solving the Heisenberg equations of motion for the bosonic modes which are given as
\begin{equation}\label{eq:HeisenbergEom}
i\pdt\mat{b_p}{\bd_{-p}}=\frac{v_s(t)\abs{p}}{2}\mat{w(t)&-g(t)}{g(t)&-w(t)}\mat{b_p}{\bd_{-p}}.
\end{equation}
This is a system of coupled differential equations each coupling the momentum modes $b_p$ and $\bd_{-p}$. Performing a time-dependent Bogoliubov transformation
$b_p=u_p(t)b_p(0)+v_p^*(t)\bd_{-p}(0)$
and an analog transformation for $\bd_{-p}$ maps the operator time dependencies to the Bogoliubov coefficients $u_p(t)$ and $v_p(t)$. Applying these transformation to the Heisenberg equations~\eqref{eq:HeisenbergEom} leads to coupled differential equations for the coefficients. Defining $\vec R_p=(u_p(t),v_p(t))^t$ these can be written as
\begin{equation}\label{eq:EomTransform0}
i\pdt \vec R_p=M_p(t)\vec R_p,\ M_p(t)\! \equiv\! \tfrac{v_s(t)\abs{p}}{2}
\mat{w(t)&\!-g(t)}{g(t)&\!-w(t)}.
\end{equation}
In general these equations cannot be solved analytically. Since the matrix $M_p(t)$ is time-dependent, a transformation $S_p$ which diagonalizes $M_p(t)$ is in general time-dependent itself and thus leads to nonadiabatic corrections. These corrections introduce again an off-diagonal coupling of the transformed variables $\vec R_p^{(1)}\equiv S^{-1}(t)\vec R_p$. To find the corrections due to this off-diagonal coupling we calculate the matrix $S_p$ such that $S^{-1}_pM_p(t)S_p\equiv M_p^{(1)}(t)$ is diagonal and transform equation~\eqref{eq:EomTransform0} accordingly. We find that the diagonal entries of the resulting matrix $M_p^{(1)}(t)$ are then given by $\pm\abs p v_s(t)$ whereas the off-diagonal couplings $\dot S_p^{-1}S_p$ are given by $-i\dot{K}(t)/2K(t)$. The latter can be neglected if
\begin{equation}\label{eq:AdiabaticityCondition}
\abs{p}\gg\left\lvert\frac{\dot{K}(t)}{2K(t)v_s(t)}\right\rvert .
\end{equation}
 For a fully adiabatic dynamics this condition has to be fulfilled for all relevant modes $p$ at all times. However, this 
 is impossible for all momentum modes since the Hamiltonian~\eqref{eq:H_Luttinger_B} is gapless. Nevertheless, bounding the timescale of changing $K(t)$, i.e., of storing the polaritons, gives rise to a momentum scale $p_\mathrm{c}$ such that all large momentum modes $\abs{p}\geq p_\mathrm{c}$ fulfill~\eqref{eq:AdiabaticityCondition} at all times. We divide the set of momentum modes into two parts $\abs p\geq p_\mathrm{c}$ and $\abs{p}<p_\mathrm{c}$ which gives rise to two spatial regimes. On small length scales corresponding to large momentum modes the correlations can adiabatically follow the change of parameters while on large length scales corresponding to small momentum modes  they stay frozen in 
the initial state. This simple argument yields a length scale $L_0\sim 1/p_\mathrm c$ marking the crossover between the two regimes. 

A better estimate for the crossover-length between adiabatic and diabatic behavior can be found as follows:
Correlations in real space propagate at the speed of sound $v_s$, i.e., turning $v_s\sim\sqrt{K(t)}$ to zero during light storage freezes the correlations. Performing the storage in finite time lets correlations propagate up to a certain length scale $\Lc$ marking a crossover between an adiabatic regime, where the correlations follow the storage to a diabatic regime which shows the initial correlations. This length scale is given by the integral over the speed of sound
\begin{equation}\label{eq:L0_time}
L_\mathrm{corr}=\int_0^\infty \mathrm d t\ v_s(t),
\end{equation}
which we will calculate for a particular time-dependence in the following.

\subsection{Special solution}

Let us now discuss a particular time-dependence of the parameters, which allows for a semi-analytic solution of the time-dependent Luttinger model. To this end, we use the approximate dependence of $K$ on the mass from equation~\eqref{eq:K_Parameter_ODB} to introduce a time-dependence of the $K$-parameter via $\Theta(t)/\Theta_0=\Omega(0)^2/\Omega(t)^2$. Furthermore, we iterate the diagonalization procedure, i.e., we find a Matrix $S_p^{(1)}$ that diagonalizes $M_p^{(1)}$. The off-diagonal corrections in this second-order adiabatic diagonalization are then given by the time derivative
\begin{equation}\label{eq:AdiabaticityCondition2}
\pm\frac{\mathrm d}{\mathrm dt}\left(\frac{\dot K}{Kv_s}\right).
\end{equation}
For a Galilean invariant system we can express the speed of sound in terms of $K$ and $m$ via $v_s=\pi\rho_0/mK$. Plugging this into~\eqref{eq:AdiabaticityCondition2} and making use of the fact that $K$ and $m$ have a unique relation, we can find a special time dependence of $K(t)$ (or equivalently $m(t)$) such that expression~\eqref{eq:AdiabaticityCondition2} vanishes for all times. This yields a differential equation for $K(t)$ which has the solution
\begin{equation}\label{eq:Kt_protocol}
K(t)=e^{-\mathrm{acosh}(t/\tau+C)},
\end{equation}
where $C=\frac{1}{2}(K_0+1/K_0)$. With this we can calculate the critical momentum $p_c$ defined by the expression on the right of the inequality~\eqref{eq:AdiabaticityCondition},
\begin{equation}
p_\mathrm c=\left\lvert\frac{\dot K(t)}{K(t)v_s(t)}\right\rvert=\frac{1}{\pi\rho_0\eta\tau}\equiv\frac{1}{L_0} =const.,
\end{equation}
which is now constant in time. We can express the crossover length $L_0=\frac{\pi^4}{90}(\rho_0 a_B^0)^5\vg(0)\ODB^0\frac{\gamma}{\abs\Delta}\tau$ in terms of the parameters $\rho_0a_B^0$ and $\OD_B^0$ describing the initial conditions.

For the maximal distance up to which correlations can build up during the storage protocol  we find 
\begin{equation}
L_\mathrm{corr}(t)=\int_0^t\mathrm ds\ v_s(s)=\frac{L_0}{2}\ln\left(\frac{K(0)}{K(t)}\right).
\end{equation}
Apart from logarithmic corrections the maximal distance is given by $L_0/2$, allowing in principle for a large correlation length $\Lc$ if $K(t)\ll K(0)$. However, for large times $t$ the $K$-parameter vanishes like $K(t)\rightarrow\frac{\tau}{2t}$, i.e., it takes long times to reach small $K$ values. This limits the maximal $L_\mathrm{corr}(t)$ in practical realizations of the protocol.

Switching off the control field such that $K$ gets tuned to zero according to~\eqref{eq:Kt_protocol}, the off-diagonal coupling of the twice transformed equations vanishes and we can integrate the differential equations to get the solution
\begin{equation}
\vec R_p^{(2)}(t)=\mat{e^{-i\xi(t)/L_0}&0}{0&e^{i\xi(t)/L_0}}\vec R_p^{(2)}(0)
\end{equation}
where $\xi(t)=\frac{1}{2\Lc(t)}\sqrt{p^2L_0^2-1}$. It is now straightforward to calculate the time evolution of the operators $\bd_p(t)$ and $b_p(t)$ by inverting the transformations. 

\subsection{Correlation functions}
%
To calculate the time evolution of the correlation functions we have to assume an initial state. The two-photon initial state we find under stationary EIT conditions is close to the two-polariton ground state. Thus we assume in the following as initial state the ground state or a low-temperature thermal state. The initial time-independent 
LL-Hamiltonian for $t\leq 0$ can be diagonalized by means of a Bogoliubov transformation which is given by the transformation $S=S_p(0)$ as defined above, yielding the Hamiltonian
\begin{equation}
\Hrm=\frac{1}{2}v_s(0)\sum_{p\neq 0}\abs{p}\gamma^\dagger_p\gamma_p.
\end{equation}
If the initial state is the ground state ($T=0$) or a thermal state ($T>0$ ) of this Hamiltonian the initial correlation functions are given by
\begin{equation}\label{eq:initial_correlations_T}
\braket{\gamma_p\gamma^\dagger_q}=\delta_{p,q}\coth\Bigl(v_s\abs{p}/2L_T\Bigr),
\end{equation}
where $L_T$ denotes the thermal length corresponding to a finite temperature $T$. 

To begin with, we consider the case $T=0$. Using the initial correlations~\eqref{eq:initial_correlations_T} in this case and transforming back to operators $\bd_p,\ b_p$ we calculate the time-dependent correlation functions, in particular the equal time density-density correlations which are in Luttinger liquid theory universally given by equation~\eqref{eq:density_correlations_luttinger}.
The time-dependent correlator $G_{\phi\phi}$ for our protocol is now given by
\begin{align}\label{eq:correlator_phiphi_quench}
G_{\phi\phi}(t)&=
\int\frac{\mathrm dp}{p}(1-\cos(pz))\braket{(\bd_p+b_{-p})(\bd_{-p}+b_p)}\nonumber\\
&\!\begin{multlined}
=K(t)\int_0^\infty\!\!\mathrm dp\ \frac{e^{-\alpha p}}{p}(1-\cos(pz))\\
\times\left[1+\frac{1-\cos(2\xi(t)/L_0)}{L_0^2p^2-1}
-\frac{\sin(2\xi(t)/L_0)}{\sqrt{L_0^2p^2-1}}\right],
\end{multlined}
\end{align}
where we introduced a cutoff $\alpha$ to treat UV divergences. The full integral can only be evaluated numerically, but we can analyze the limiting cases for small and large momenta $p$ giving us asymptotic results for large and small distances $z$, resp.  Since $L_\mathrm{corr}(t)$ is growing only logarithmically with $K(t)$ we assume $L_\mathrm{corr}(t)\gtrapprox L_0$.  For $pL_0\gg 1$ the terms $\cos(2\xi/L_0)$ and $\sin(2\xi/L_0)$ oscillate quickly and thus average to zero. The remaining integral can be calculated analytically and gives the fully adiabatic correlations
\begin{equation}
G_{\phi\phi}=-K(t)\ln\left(\frac{\alpha}{z}\right)+const.
\end{equation}
leading to a power law in the density-density correlations as in the initial ground state but now decaying with the much smaller exponent $2 K(t)\ll 1$, thus indicating a quasi-crystalline order. For $pL_0\ll 1$, on the other hand, $\xi(t)$ becomes purely imaginary and approximates to $\ln(K(0)/K(t))$ such that the last line in~\eqref{eq:correlator_phiphi_quench} approximates to $\frac{K(0)}{K(t)}$ cancelling the prefactor $K(t)$ and indicating a power law decay with the {\em initial} exponent $2K(0)$. These limits agree with the crossover from an adiabatic to a diabatic regime which we expected from the gaplessness of the model.

\begin{figure}
\vspace*{\baselineskip}
 \centering
\includegraphics[width=\columnwidth]{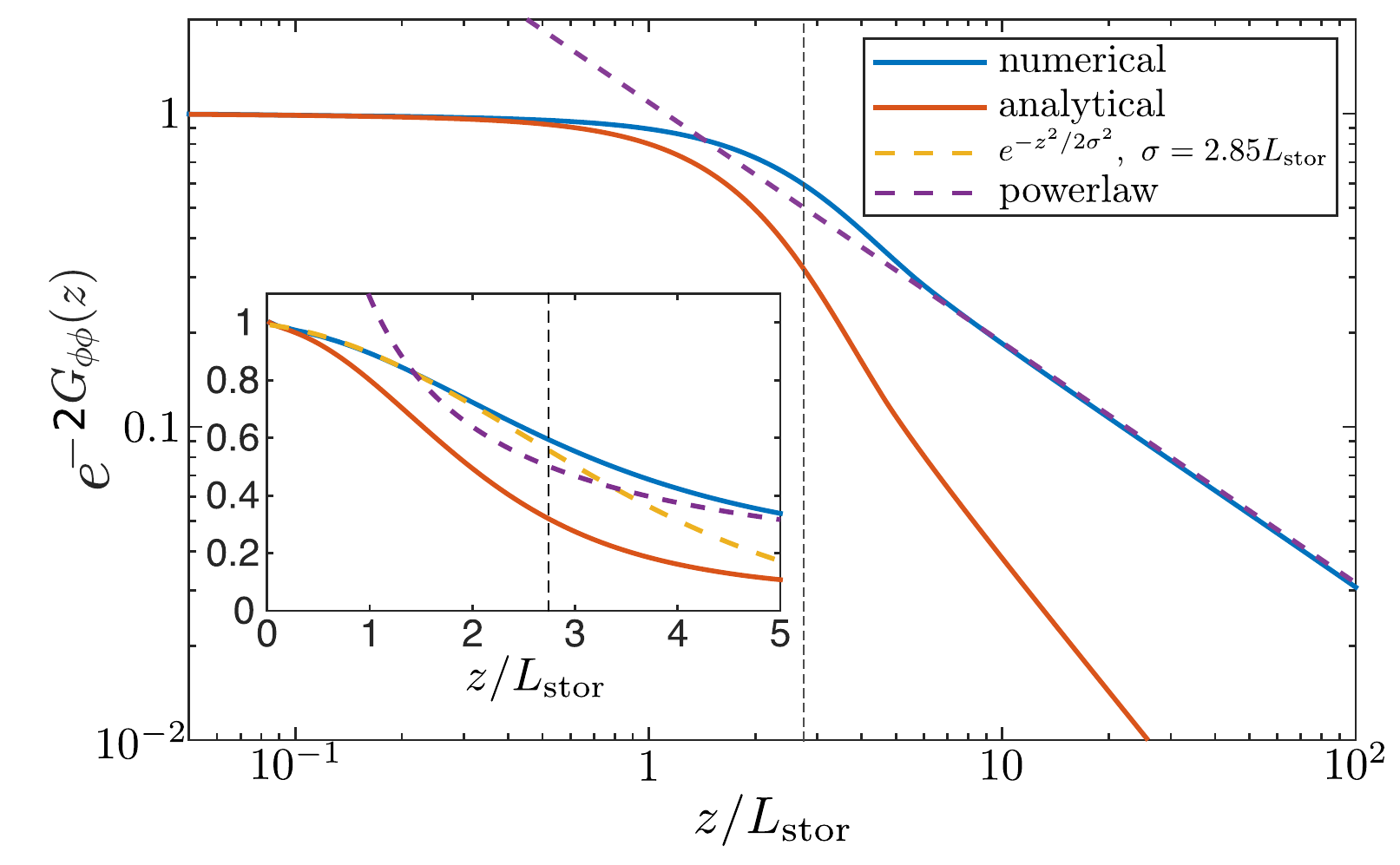}
 \caption{%
Spatial envelope $\exp\{-2G_{\phi\phi}(z)\}$ of the CDW correlation functions for storage the storage protocol~\eqref{eq:Kt_protocol}. Comparison of full numerical integration of the time evolution using interpolated $K(\Theta)$
  to the analytical result~\eqref{eq:correlator_phiphi_quench} using formula~\eqref{eq:K_Parameter_ODB}, where we fix the initial $\Theta_0$ for both results. The correlation function shows a crossover from an adiabatic regime, where the spatial decay fits well to a Gaussian function, to a diabatic regime, showing a power law with initial exponent $2K_f$. The dashed vertical lines indicate the distance $z=L_\mathrm{corr}$. \textit{Inset:\ } Plotting the correlations on linear scale show that for small distances the correlation functions are well described by a Gaussian function with the width $\sigma=L_\mathrm{corr}$.
  }%
 \label{fig:Correlations_Quench}
\end{figure}

We calculate the density-density correlation function~\eqref{eq:correlator_phiphi_quench} after the storage numerically and compare it to a full numerical integration of~\eqref{eq:EomTransform0} using an interpolated $K(\Theta)$ according to the DMRG results in Fig.~\ref{fig:K_ODB}. In Fig.~\ref{fig:Correlations_Quench}
we show space-dependent amplitude  $e^{-2G_{\phi\phi}(z)}$ of the density-density correlations~\eqref{eq:density_correlations_luttinger} of both results, for the same intial $\Theta_0$.

We plot the distance in units of the length $L_\mathrm{stor}$, defined by the distance a wave-packet propagates during the storage protocol, which we will introduce below. 

The correlations show the expected crossover at length scale $L_\mathrm{corr}$ from an adiabatic to a diabatic regime, where asymptotically the correlation function decays as a power law with initial exponent $K_0$. Note that we choose the same initial $\Theta_0$ for both, analytical and numerical calculation, which yields different initial $K_0$, since the analytic calculations use the approximate relation between $K$ and $\Theta$. As can be seen from the inset in Fig.~\ref{fig:Correlations_Quench} the adiabatic spatial regime is well described by a Gaussian profile $\exp\{-z^2/2\sigma^2\}$. From fitting this profile to the calculated correlations we can extract the full width at half maximum $\sigma$ of the Gaussian giving the value where the amplitude of density-density correlations has dropped to $1/2$, yielding a length scale for the crossover from adiabatic to diabatic regime corresponding to $L_\mathrm{corr}$.

\subsection{Limitations.}
In experimental realizations of the proposed scheme only finite medium lengths are available. Thus the time scale $\tau$ of the protocol has to be limited such that the distance the polaritons travel during storage is less than the medium length.
During the protocol~\eqref{eq:Kt_protocol} the polaritons propagate the distance 
\begin{equation}
L_\mathrm{stor}=\int\mathrm ds\ \vg(s) \leq \vg(0)\Theta_0K_0\tau.
\end{equation}
As for a finite system this scale is ultimately limited, we compare $L_\mathrm{stor}$ to the correlation length scale $L_\mathrm{corr}$. The timescale $\tau$ drops out and we get the best possible $\Lc$ for a given $L_\mathrm{stor}$ 
as
\begin{equation}
\frac{L_\mathrm{corr}}{L_\mathrm{stor}}=\frac{\pi^4}{45}\frac{\rho_0 L_{\mathrm abs}}{K_0}\frac{\abs\Delta}{\gamma}\ln\left(\frac{K_0}{K_\mathrm f}\right),
\end{equation}
where the initial $K_0$ is also dependent on $\rho_0a_B$ and $\ODB$. A feasible $ K_0 $ can be read off Fig.~\ref{fig:K_ODB} showing that it is possible to get strong correlation functions on the order of the medium length for sufficiently small final $K_\mathrm f$. 

\subsection{Corrections}

In the following we consider corrections to the results due to deviations of the initial state from the ground state and additional losses during storage due to nonadiabatic couplings.

\begin{figure}
 \centering
\includegraphics[width=\columnwidth]{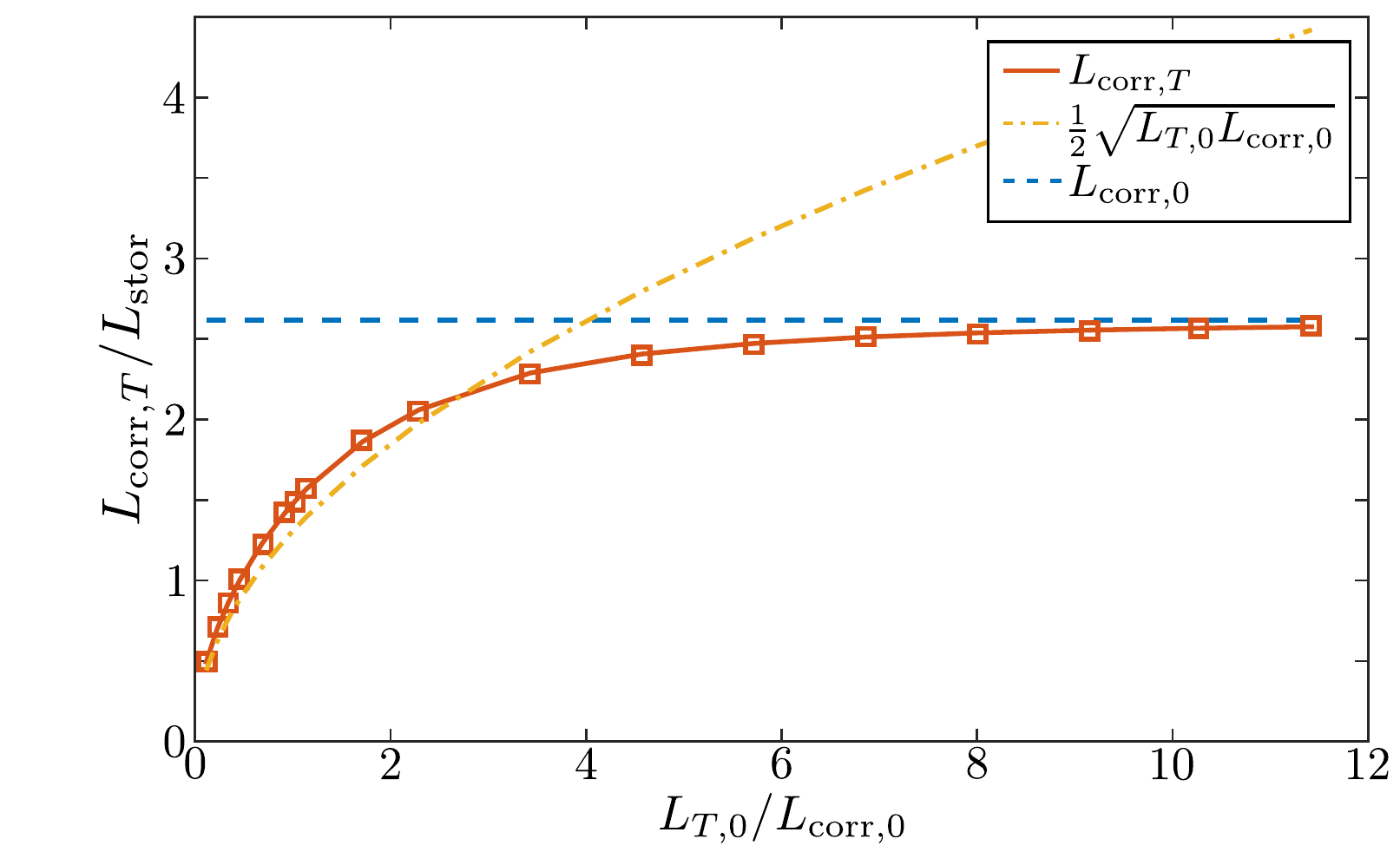}
 \caption{%
Correlation length scales after storage for a thermal initial state. The red squares show the thermal length scale from a Gaussian fit to the correlation function $\exp\{-2G_{\phi\phi}(z)\}$, which we compare to the geometric mean of $L_{T}$ and $L_\mathrm{corr}$ for initial ground state. As can be seen, for small thermal length scales the resulting correlation length scale is given by this geometric mean and saturates asymptotically to $L_\mathrm{corr}$ for $L_T\gg L_\mathrm{corr}$.
  }%
 \label{fig:LT}
\end{figure}

\begin{figure}
 \centering
\includegraphics[width=\columnwidth]{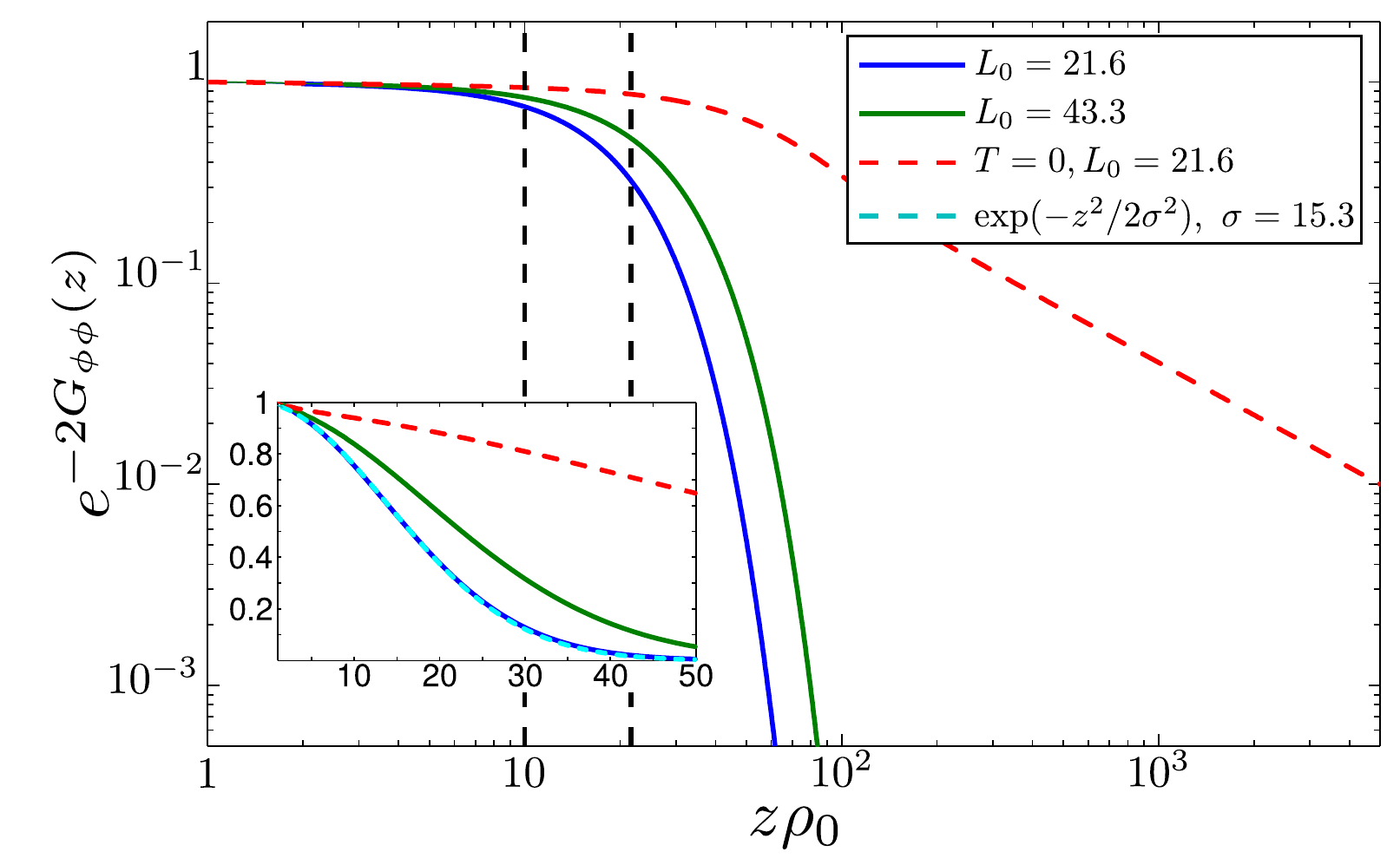}
 \caption{%
Correlation function $\exp\{-2G_{\phi\phi}(z)\}$ for a fixed finite initial Temperature $T_0$ and two different crossover length scales. The resulting crossover increases with $L_0$. Vertical lines correspond to the length scales $L_0=21.6$ and
  }%
 \label{fig:Correlations_Quench_T}
\end{figure}

\paragraph*{Initial excitations.}As we have shown numerically the initial state created under stationary propagation of a two-photon pulse into a Rydberg medium is close to the ground-state.  As the Rydberg polaritons can only be excited inside the EIT window~\eqref{eq:EIT_window} we can estimate the maximum allowed momentum fluctuations by $\abs k\leq\abs{k_\mathrm{max}}=\Omega_e^2/ \abs{\Gamma}c$. Excitations with a kinetic energy corresponding to $k>k_\mathrm{max}$ couple to bright-state polariton degrees of freedom which propagate almost at the vacuum speed of light, or, in the case of near single-photon resonance, are absorbed and thus quickly disappear. Therefore, modeling excitations of the initial state within the EIT window by a finite initial temperature, we can estimate an upper bound for this temperature by $k_\mathrm BT_0\leq (\Omega_0^2+g^2\sqrt n)/2m\abs\Gamma c$, yielding a thermal length scale~\cite{Cazalilla2004} of 
($k_B=1$)
\begin{equation}
L_T\geq\frac{v_sK}{\pi T}=2\rho_0a_B\frac{\abs\Delta}{\gamma}\frac{1}{\ODB}.
\end{equation}
We then take the thermal excitations into account in the initial correlation functions~\eqref{eq:initial_correlations_T} of the time-dependent Luttinger calculation. The thermal length scale $L_T$  marks a crossover of the initial state from power law decay of the spatial correlation functions to an exponential decay~\cite{Cazalilla2004}. We again calculate the time evolution of the density-density correlation functions during the storage and plot the resulting functions in Fig.~\ref{fig:Correlations_Quench_T}.  We find that while the correlation function of the thermal initial state decays exponentially for $z\gg L_T$, the correlation function after storage shows adiabatic following for a region larger $L_T$. In particular by fitting a Gaussian to the correlation function and extracting the crossover length scale, we find that after storage instead of the $T=0$-correlation length $L_\mathrm{corr}$ the relevant length scale is $L_{\mathrm{corr},T}$ 
which is
\begin{equation}
L_{\mathrm{corr},T}=\frac{1}{2}\sqrt{L_\mathrm{corr}L_T}.
\end{equation}
Numerically we find that this condition holds as long as $L_T\lesssim 2 L_\mathrm{corr}$, while for large initial thermal length, $L_{\mathrm{corr},T}$ asymptotically approaches the zero-temperature result $L_\mathrm{corr}$. This is shown in Fig.~\ref{fig:LT}

From this we conclude that if the storage can be done sufficiently slow, if $L_\mathrm{stor}$ is sufficiently large, a finite range Wigner crystal can be created despite initial thermal excitations or imperfections.

\paragraph*{Non-adiabaticity effects.}
In a \textit{cw} experiment the validity of our model is guaranteed as long as the initial pulse spectral width fits well inside the EIT window defined by condition~\eqref{eq:EIT_window}. For storage, i.e., a time-dependent control field, however, we have to modify condition~\eqref{eq:EIT_window} and additionally consider $ \dot{\theta}\ll\sin\theta\cos\theta\abs{\Gamma_\mathrm{eff}} $. This restricts the time scales on which the input pulse can be stored. For our specific protocol the nonadiabatic coupling $\sim\dot\theta$ is bounded by its value at $t=0$.  We find
\begin{equation}
\frac{2K_0}{(K_0^2-1)^2}\ll\frac{\gamma}{\abs\Gamma}\frac{c\tau}{\La}.
\end{equation}
Solving for $\tau$ and using the fact that the maximal allowed $\tau$ is bounded by the condition that the polaritons may maximally propagate the system length during storage, i.e., $L_\mathrm{stor}\leq L$ we can write this condition as lower bound on the total optical depth of the system,
\begin{equation}
\mathrm{OD}\gg\frac{90\abs\Gamma\vg(0)}{\pi^4\gamma c}\frac{1}{1-K_0^2},
\end{equation}
which obviously diverges for $K_0\rightarrow 1$. However, as we can read off from Fig.~\ref{fig:K_ODB} an initial $K_0<0.8$ is experimentally feasible for stationary EIT conditions. Assuming $\abs\Gamma \vg (0)/\gamma c\lesssim 1$ we find $\mathrm{OD}\gg 3$, which is easily achievable in current experimental realizations~\cite{Firstenberg2013}.

\section*{Summary}
In summary, we derived an effective model for the interaction of photons in a gas of Rydberg atoms under conditions of electromagnetically induced transparency (EIT) in particular in the regime of large optical depth per blockade distance, $\ODB\gg 1$. We showed that under paraxial propagation conditions and for sufficiently low densities of excitations the system can be described by an effective model of a single species of quasi-particles, called Rydberg polaritons, which becomes one-dimensional, if the transverse beam diameter is less than the EIT blockade radius.  For sufficiently large inter-particle distances Rydberg polaritons behave as massive Schr\"odinger particles with repulsive van der Waals type interactions. For shorter distances there is a coupling of dark- to fast propagating bright-state polaritons, which gives rise to an effective loss-mechanism for Rydberg polaritons\cite{Firstenberg2013}. For an off-resonant excitation scheme with finite single-photon detuning also bound two-particle states exists. As will be discussed in detail elsewhere~\cite{UnanyanPrep} these states can not be excited for large $\ODB$ from an initial light field and thus were not considered here. 
We derived conditions where the losses in the effective Rydberg polariton model are negligible and we can use an effective Hamiltonian. The ground-state properties of this Hamiltonian were analyzed using numerical DMRG simulations and in terms of a Luttinger liquid model. We showed that the regime of strong interactions, quantified by the ratio of interaction to kinetic energy $\Theta \gg 1$ is very difficult to reach under stationary EIT conditions. Increasing the strength of the Rydberg interaction leads to an increase of the EIT
blockade distance, which prevents to reach sufficiently large polariton densities to enter the
strong-interaction regime.  However, making use of a dynamical slow-down of Rydberg polaritons  while propagating inside the medium or a storage protocol of polaritons in stationary spin excitations allows to decrease the kinetic energy by
increasing their effective mass without reducing the quasi-particle density. 
In this way it is possible to generate a quasi-crystalline or a charge-density wave (CDW) 
state of stored photons, which is a highly non-classical state consisting of an ordered string of single-photon wave-packets. 
This state can be observed by either nonadiabatic release of the stored excitations into a train of single photons or by direct imaging
of the Rydberg ensemble as in~\cite{Guenter2013}.
We analyzed this storage in terms of a time-dependent Luttinger liquid model and showed that the gapless model leads to a spatial crossover in the CDW correlations between an adiabatic regime, exhibiting strong spatial correlations and a diabatic regime, where the correlations show the initial power-law decay.

We thank J. Otterbach, S. Whitlock, M. Weidem\"uller and H. P. B\"uchler for valuable discussions.  The financial support of the DFG through SFB-TR49 is gratefully acknowledged.



\end{document}